\documentclass[review, 3p]{elsarticle}

\usepackage{amsmath, amsfonts}
\usepackage{adjustbox}
\usepackage{algorithm}
\usepackage{array}
\usepackage[caption=false,font=normalsize,labelfont=sf,textfont=sf]{subfig}
\usepackage{textcomp}
\usepackage{multirow}
\usepackage{stfloats}
\usepackage{pifont}
\usepackage{url}
\usepackage{hyperref}
\usepackage{verbatim}
\usepackage{graphicx}
\usepackage{tabularray}
\usepackage{threeparttable}
\usepackage{booktabs}
\usepackage{todonotes}
\usepackage{longtable}
\usepackage[font=small,labelfont=bf]{caption}
\usepackage{makecell}
\usepackage{ragged2e}
\usepackage{soul}
\soulregister\cite7

\journal{Neurocomputing}

\begin{document}

\begin{frontmatter}



\title{From Data Heterogeneity to Convergence: A Data-Centric Review of Federated Learning}
\author[1]{Huong Nguyen}\ead{huong.nguyen@oulu.fi}
\author[2]{Micka{\"e}l Bettinelli}\ead{mickael.bettinelli@univ-smb.fr}
\author[1]{Amirhossein Ghaffari}\ead{amirhossein.ghaffari@oulu.fi}
\author[2]{Alexandre Benoit}\ead{alexandre.benoit@univ-smb.fr}
\author[3]{Hong-Tri Nguyen}\ead{hong-tri.nguyen@aalto.fi}
\author[1]{Susanna Pirttikangas}\ead{susanna.pirttikangas@oulu.fi}
\author[1]{Lauri Lov{\'e}n\corref{cor1}}\ead{lauri.loven@oulu.fi}

\affiliation[1]{organization={Future Computing Group, University of Oulu},
            country={Finland}}
\affiliation[2]{organization={LISTIC, University Savoie Mont Blanc},
            country={France}}
\affiliation[3]{organization={AaltoSEA Group, Aalto University},
            country={Finland}}
\cortext[cor1]{Corresponding author}

\begin{abstract}
Federated Learning (FL) has emerged as a promising solution for data hunger in centralized learning. This paradigm enables privacy with multiple clients to train a shared-task model collaboratively without exposing their local data. While being a key component in any learning system, data is also a primary source of vulnerabilities and challenges, and a major determinant of a stable and well-converged training. Existing FL reviews describe general foundations, security practices, opportunities, challenges, and applications, without delving into diverse aspects of data and considering problems from the data perspective. They rarely provide a data-lens synthesis that links concrete data properties, split protocols, and defenses to convergence speed and stability. This survey fills that gap with three advances. First, we analyze ``non-IID'' into measurable traits and rank their influence on convergence as strong, medium, or light, explaining the mechanisms behind each and reconciling evidence across images, texts, and graphs. Second, we connect experimental splitting practices to the real phenomena they emulate, expose the artifacts they introduce, and show how those artifacts affect target accuracy. Third, we analyze how data-related vulnerabilities and their proposed defenses affect convergence, reporting performance under clean and adversarial conditions to make the convergence-robustness trade-off explicit. To our knowledge, this is the first survey to provide a complete understanding of data-related challenges that govern FL. With clear takeaways distilled for each concern, our work serves as actionable guidance, helping practitioners design their system with predictable convergence and stability.
\end{abstract}


\end{frontmatter}



\section{Introduction}
Recent advances in artificial intelligence (AI) have led to increasingly sophisticated data modeling paradigms, including graph-based learning, low-rank structured optimization, and multi-view representation learning, which exploit intrinsic data structures to enhance predictive performance \cite{yan2020deep, yan2020depth}. These approaches have demonstrated that modeling relational structure, feature heterogeneity, and multi-view consistency can significantly influence optimization behavior and downstream accuracy in centralized settings. However, training such models typically requires large volumes of high-quality data, which are often fragmented across multiple devices or institutions. This poses fundamental challenges in centralized learning where a single entity frequently lacks sufficient data for robust model training. As a remedy, Federated Learning (FL)~\cite{mcmahan2017communication} enables multiple participants to collaboratively train a shared model across distributed environments without centralizing their datasets, thereby addressing data scarcity in centralized ML and finding success in real-world applications~\cite{nguyen2021federated-domain}.

Despite its promise, FL's practical behavior is driven more by data decisions, the traits of the source data, how they are split across clients, and the side effects introduced by defense mechanisms. Yet, most existing surveys adopt a method- or system-centric lens \cite{nguyen2021federated-domain, zhang2021survey, wen2023survey, liu2022distributed, aledhari2020federated, abdulrahman2020survey}, and even more specific ones like security- and heterogeneity-oriented overviews tend to treat these issues broadly rather than analyzing their concrete trade-offs~\cite{zhu2021federated, mothukuri2021survey}. To address this gap, we present a data-centric examination that considers data as a primary driver of convergence speed, stability, and final accuracy, organizing the survey around how (i) data traits, (ii) splitting protocols, and (iii) data-related defenses shape FL outcomes separately.

To be more specific, while established results show that more data and fewer classes generally improve performance and accelerate convergence \cite{hestness2017deep, sun2017revisiting, zhu2012we}, the literature offers little systematic assessment of other intrinsic data traits--e.g., distribution skew, provenance (cross-device vs. cross-silo), or feature dimensionality--and, critically, how strongly each trait influences the convergence. We compile evidence across benchmark sources in prior experimental works to discuss and then provide simple design guidance for dataset design based on traits' impacts. 

In another regard, most of these widely used data were not designed for FL and therefore require intentional partitioning to emulate real deployments. The choice of split protocol directly determines how similar client updates are, the convergence rate, and how fast the model hits a target accuracy. We thus explore the associated challenges and the common splitting methods, as well as how they affect convergence.

Finally, often the main reason for shifting from centralized to federated training is FL's privacy-preserving distributed architecture. However, the distributed architecture motivates attackers to breach these guarantees and disrupt learning. Because data is central to any ML system, it becomes a prime target for malicious activities. Accordingly, we systematically examine vulnerabilities from a data-centric perspective. To enhance clarity within this vulnerability section, we provide two summary tables. The first is structured around "The Five Ws and How" questions (What, Who, When, Where, Why, and How) to offer a holistic overview of each attack. The second table complements this by summarizing defense mechanisms for each data-related threat, reporting core idea, clean utility (no-attack), utility under attack, and our brief notes on convergence or overhead. This makes the protection–performance trade-off explicit and highlights when defenses preserve near-clean accuracy versus when robustness gains incur some material costs.

Guided by this perspective, the survey is organized around four research questions:

\begin{itemize}
    \item \textbf{RQ1:} What features of the data source affect convergence speed and training performance? (Section~\ref{sec:opensrc})
    \item \textbf{RQ2:} What factors prevent simulation from accurately reflecting real-world data? (Section~\ref{sec:spliting})
    \item \textbf{RQ3:} How are datasets typically split for federated settings in prior experiments? (Section~\ref{sec:spliting})
    \item \textbf{RQ4:} What are the convergence-privacy trade-offs of common defense methods? (Section~\ref{sec:security})
\end{itemize}

By addressing these questions, this review aims to provide a comprehensive guide to understanding and overcoming the data-related challenges in FL, paving the way for more robust and comprehensive learning systems. All sections throughout this study are approached in a data-centric manner, considering data-driven perspectives. 

The remainder of this paper is organized according to our RQs as follows:
Section~\ref{sec:method} describes the survey methodology, clarifying how we found the articles. 
Section~\ref{sec:opensrc} systematically analyzes the data characteristics and their impact on convergence behavior.
Section~\ref{sec:spliting} brings up the concern on how data should be split for federated settings.
Section~\ref{sec:security} focuses on the common data-related vulnerabilities and their defense method in FL systems. 
Section~\ref{sec:discuss} highlights the major findings, implications from our survey, and directions for future work.
Finally, Section~\ref{sec:conclusion} concludes this work.

\section{Survey methodology and scope declaration}
\label{sec:method}

\subsection{Methodology}
This paper employed a structured approach methodology to identify relevant research. 
\begin{itemize}
    \item Our primary search strategy involved using keyword combinations related to the specific topic under investigation, coupled with "federated learning", within the Google Scholar search engine. For example, "federated learning" and "attack" are put together in the search bar. 
    \item We first prioritized articles based on citation count and publication date, favoring highly cited and recent works to capture the current state-of-the-art. 
    \item Secondary criteria for article selection included the reputation and ranking of the publication venue (conference or journal) based on quartiles (Q1, Q2).
    \item For the specific analysis of how data splitting affects model convergence, we restrict the evidence table to publications from A to A*-ranked venues (by CORE rank\footnote{\url{https://portal.core.edu.au/conf-ranks/}}) such as AAAI, ECML, ICLR, ICCV, CVPR, CIKM, WWW, ICML, KDD, and NeurIPS (2020–2026) to capture recent, high-quality works in the field. This choice is intended to support a scoped comparative synthesis rather than an exhaustive cross-venue inventory. That said, we also discuss representative relevant studies from other venues in the text of this section for completeness.
    \item It is noted that while reducing reference to unofficial works, our survey also incorporated pre-print articles from arXiv that had been accepted for publication in a peer-reviewed scientific venue but had not yet been formally published by the respective publisher.
\end{itemize}

This approach ensured the inclusion of cutting-edge research while maintaining a focus on peer-reviewed and high-quality contributions.

\subsection{Scope}
We define our data-centric perspective in two primary lines: data as a determining factor, and data as a reason. Accordingly, this survey restricts coverage to aspects of federated learning that originate from data, where the cause, analysis, and conclusions are driven mainly by data properties and data processing choices.

To clarify, we acknowledge that environment and system-level constraints can influence how data handling choices are implemented and may indirectly affect training behavior and convergence. That said, these side factors are outside the scope of the survey and will not be further discussed. This scope boundary keeps the paper strictly data-centric, hence, ensures a focused taxonomy and deeper synthesis of data-originated issues without expanding into a general survey.

\section{Data sources}
\label{sec:opensrc}

Prior FL works report markedly different convergence speeds and asymptotic accuracies across datasets, even when the model architecture, optimizer, and hyperparameters are held fixed. A central hypothesis is that these discrepancies stem from data-source traits rather than algorithmic choices. Sources differ not only in modality (image, text, graphs) but also in their statistical structure (e.g., label distribution skew, covariate diversity), task geometry (intra-class variability (ICV) and inter-class separation (ICS)), output space (number and granularity of classes), effective dimensionality (resolution or feature size), and provenance style (cross-device vs. cross-silo-like partitions). These traits alter gradient variance and bias across clients, reshape the local loss landscapes, and change how quickly global aggregation reduces disagreement, thereby directly affecting rounds-to-stability and final utility. 

This motivates the following question--one for which, to date, there is no consensus or official answer: \textit{``Which intrinsic traits of a data source speed up or slow down convergence--and how strongly do they influence final accuracy--when training conditions (model and hyperparameters) are controlled?''}

Answering this question matters for several reasons.
First, practitioners must set model capacity, client participation, local steps, learning-rate schedules, and preprocessing (e.g., augmentation, deduplication, label cleaning) before training; knowing which intrinsic data traits—such as label-space granularity, intra/inter-class variability, covariate diversity, and provenance (cross-device vs. cross-silo)—tend to accelerate or slow convergence gives concrete guidance for those choices (e.g., when to favor smaller models and higher client fractions versus stronger regularization and longer warm-ups). Second, there is currently no agreed-upon mapping from specific traits to expected convergence behavior; evidence is scattered across papers using different models and setups, so insights remain anecdotal. Providing a clear trait-behavior map turns that scattered evidence into actionable expectations: it helps plan training budgets, pick sensible stopping criteria and target accuracies, and select or curate datasets whose traits match deployment needs. Third, a trait-based view improves comparability and reproducibility: it explains why two sources with similar size behave differently, and it offers simple, reportable descriptors that future studies can include so their results translate to new settings.

To answer this, we (i) assemble a benchmark-oriented table that records per-source traits, (ii) annotate each source with compact convergence evidence (\textit{metric@round} if applicable) along with statistical values (median, min, max), and (iii) synthesize how each trait influences model performance. This design isolates data-driven factors from training-protocol factors, enabling a principled assessment of which traits are the dominant drivers of convergence behavior in FL.

\subsection{Benchmark sources and their traits}

Benchmark datasets play a critical role in evaluating FL methods, but their use introduces dataset-specific challenges that can significantly influence the validity and comparability of results. Table \ref{tab:src-benchmark-characteristics} summarizes commonly used FL sources across images (MNIST, Fashion-MNIST, EMNIST, SVHN, CIFAR-10/100, Tiny-ImageNet, FEMNIST), text (Shakespeare, Sentiment140), and graphs (Cora, CiteSeer, PubMed). For each dataset, we profile provenance, input size, number of samples, number of classes, native label skew, ICV, ICS, and graph homophily where applicable. These attributes determine how hard the learning problem is, how sensitive it is to non‑IID (non-independent and identically distributed) partitions, and how well results transfer to more realistic settings.

\begin{table}[htbp!]
\centering
\caption{Standard Benchmark Sources: Data Characteristics.}
\label{tab:src-benchmark-characteristics}
\scriptsize
\setlength{\tabcolsep}{2pt}
\renewcommand{\arraystretch}{1.05}
\begin{adjustbox}{max size={\linewidth}{\textheight}}
\begin{tabular}{|
  >{\centering\arraybackslash}p{0.115\textwidth}| 
  >{\centering\arraybackslash}p{0.115\textwidth}|  
  >{\centering\arraybackslash}p{0.105\textwidth}|    
  >{\centering\arraybackslash}p{0.08\textwidth}|    
  >{\centering\arraybackslash}p{0.07\textwidth}|   
  >{\centering\arraybackslash}p{0.065\textwidth}|   
  >{\centering\arraybackslash}p{0.075\textwidth}|  
  >{\centering\arraybackslash}p{0.075\textwidth}|   
  >{\centering\arraybackslash}p{0.075\textwidth}|    
  >{\centering\arraybackslash}p{0.075\textwidth}|}
\hline
{\centering\textbf{Dataset}} &
{\centering\textbf{About}} &
{\centering\textbf{Provenance (Cross-X)}} &
{\centering\textbf{\# Dimension}} &
{\centering\textbf{\# Samples}} &
{\centering\textbf{\# Classes}} &
{\centering\textbf{Native label skew}} &
{\centering\textbf{ICV}} &
{\centering\textbf{ICS}} &
{\centering\textbf{Graph homophily}} \\
\hline

\textbf{MNIST} & Handwritten digits (images) & Generic &
$1\times 28\times 28$ & 70{,}000 & 10 & \ding{55} & Low & High & - \\
\hline
\textbf{Fashion-MNIST} & Fashion (images) & Generic &
$1\times 28\times 28$ & 70{,}000 & 10 & \ding{55} & Med & Med & - \\
\hline
\textbf{EMNIST} & Handwritten letters and digits (images) & Generic &
$1\times 28\times 28$ & 814{,}255 & 26--62 (split depend) & $\triangle$ (split depend) & Med & Med & - \\
\hline
\textbf{SVHN} & Street-view digits (images) & Generic &
$3\times 32\times 32$ & 630{,}420 & 10 & $\triangle$ (moderate) & Med-High & Med & - \\
\hline
\textbf{CIFAR-10} & Natural images & Generic &
$3\times 32\times 32$ & 60{,}000 & 10 & \ding{55} & Med & Med & - \\
\hline
\textbf{CIFAR-100} & Natural images & Generic &
$3\times 32\times 32$ & 60{,}000 & 100 & \ding{55} & High & Low & - \\
\hline
\textbf{Tiny-ImageNet} & Natural images (Subset of ImageNet) & Generic &
$3\times 64\times 64$ & 120{,}000 & 200 & \ding{55} & High & Low & - \\
\hline
\textbf{FEMNIST (LEAF)} & Federated version of EMNIST (images) & Cross-device (per writer) &
$1\times 28\times 28$ & 805{,}263 & 62 & \ding{51} & Med & Med & - \\
\hline
\textbf{Shakespeare (LEAF)} & Next character prediction (text) & Cross-device (per role) &
Seq.\ len$=80$ tokens & 4{,}226{,}158 & --- & \ding{51} & High & - & - \\
\hline
\textbf{Sentiment140 (LEAF)} & Tweets (text) & Cross-device (by user) &
Seq.\ len / BoW & 1{,}600{,}498 & 2 & \ding{51} & High (short, sparse) & High & - \\
\hline
\textbf{Synthetic (LEAF)} & Generated & Cross-device (cfg.) &
cfg. & cfg. & cfg. & cfg. & cfg. & cfg. & - \\
\hline
\textbf{Cora} & Citation graph & Cross-silo (partitioned into subgraphs in FL) &
1433-d BoW & 2{,}708 & 7 & - & Med & Med--High & High \\
\hline
\textbf{CiteSeer} & Citation graph & Cross-silo (partitioned into subgraphs in FL) &
3703-d BoW & 3{,}312 & 6 & - & Med & Med & Med \\
\hline
\textbf{PubMed} & Citation graph (diabetes-related) & Cross-silo (partitioned into subgraphs in FL) &
500-d TF-IDF & 19{,}717 & 3 & - & Med & High & Med--High \\
\hline

\end{tabular}
\end{adjustbox}
\end{table}

\subsubsection{Image}
Widely used image benchmarks such as MNIST and its variants provide convenient testbeds for non-IID data splits, yet they also risk oversimplifying the evaluation landscape. In particular, studies often evaluate MNIST under pathological non-IID settings (e.g., two labels per client), leading to extreme label imbalance and client drift \cite{parsons2023mobilizing}. The dataset is also highly sensitive to mislabeled clients \cite{chen2023space}, and baseline aggregators are very vulnerable to backdoor attacks unless defenses rely on a clean server-side validation set \cite{jia2023fedgame}. In general, with low-dimensional, balanced, low ICV, and high ICS, MNIST can overstate generalization relative to harder image datasets such as CIFAR‑10, which requires deeper models and more communication \cite{chen2023space, parsons2023mobilizing}. 

At a larger scale, CIFAR-10 and CIFAR-100 expose additional challenges. CIFAR‑10 has medium ICV and medium ICS under standard preprocessing. However, under non‑IID label splits, accuracy drops, and communication efficiency becomes a bottleneck in this dataset. CIFAR‑100 adds many visually similar images across classes under the same number of samples, which drives ICV higher and compresses pairwise margins (ICS lower). It is more sensitive to class imbalance and yields higher variance across clients \cite{sun2023understanding, wang2024fednlr, xu2024cooperative}. 

SVHN collects street‑view digits with cluttered backgrounds. The background variation in this dataset raises ICV to medium-high and reduces ICS to medium, and class frequencies are often moderately skewed in practice. 

EMNIST expands MNIST to letters and digits. Notably, the number of classes and the level of skewness depend on the official split of the dataset user. EMNIST's ICV and ICS are typically in the medium range. 

FEMNIST, derived from EMNIST, introduces extreme client heterogeneity because users contribute widely varying amounts of data with distinct handwriting styles. This imbalance amplifies personalization demands and destabilizes convergence across clients, posing more substantial challenges than MNIST for cross-device FL \cite{yang2023dynamic, wang2023fluid}.

Tiny‑ImageNet increases both resolution and the number of classes. The higher class count and visual diversity raise ICV and reduce ICS further, making it a more demanding testbed than CIFAR‑10 for the same compute budget.

\subsubsection{Text}
FL studies also rely on textual benchmarks to capture the variability of human language. Shakespeare (next‑character prediction) is cross‑device and organized by role or speaker. Per‑user text is narrow in style and topic, label distributions are highly skewed, and short sequences increase ICV. Notably, because next-character prediction is a context-conditioned task, the notion of a single, context-agnostic margin between classes does not apply: class ‘clusters’ shift with the preceding text, so the ICS cannot be well defined in this dataset. 

Sentiment140 groups tweets by user from Twitter. It has two labels but substantial user‑level skew, short and noisy texts, and evolving vocabulary. These characteristics raise ICV, make local models prone to overfitting, and hinder cross-client generalization, highlighting the difficulty of personalization in large-scale text-based FL \cite{crawshaw2023federated, shi2023prior}.

Within the LEAF framework, the Synthetic dataset is fully configurable and well-suited to ablation studies that isolate the impact of a single data trait. Researchers can dial in extreme label skew, client heterogeneity, or unbalanced sample counts with simple, interpretable controls. Because it enables clean and reproducible non-IID scenarios across tasks, it has become a common choice in the FL community \cite{panchal2023flash, wan2025energy}

\subsubsection{Graph}
Citation-graph benchmarks such as PubMed, Cora, and CiteSeer are commonly used for federated GNN evaluation; however, they present unique challenges. Structural non-IIDness in these graphs makes stealthy backdoor triggers difficult to detect \cite{wan2025energy}, and many defense methods unrealistically assume access to a clean validation set. Under subgraph-FL, as shown by Kim et al., local models are prone to overfitting, suffer from missing or unseen classes, and tend to concentrate reliable knowledge in head-degree or head-class nodes, limiting generalization across clients \cite{kim2025subgraph}. Moreover, enforcing privacy by removing cross-client edges significantly degrades performance, underscoring the fundamental trade-off between graph integrity and privacy in federated graph learning \cite{kim2025subgraph, wan2025energy}.

\subsection{Data traits and their influence}
This subsection synthesizes trait-level effects on FL convergence (Table \ref{tab:traits-convergence}) by combining the dataset characteristics in Table~\ref{tab:src-benchmark-characteristics} with the cross-paper outcomes in Table~\ref{tab:src-benchmark-convergence}. We classify each trait by its observed influence on convergence speed, stability,  and final accuracy, and from that identify the most influential factors.

We interpret the convergence profiles in Table~\ref{tab:src-benchmark-convergence} via two data-dependent factors that commonly appear in convergence analyses of prior canonical works~\cite{li2019convergence, li2020federated, karimireddy2020scaffold, stich2018local}: (i) within-client stochastic gradient noise, refer to gradient variance induced by finite local data and mini-batching, and (ii) between-client heterogeneity, which leads to divergent objectives and differences in update directions across clients. From the data view, these terms are reflected in observable dataset traits. Specifically, more classes, higher input dimensionality, higher ICV, lower ICS, and sparse or ambiguous signals typically produce noisier and less consistent local updates, which increase gradient variability and can slow or destabilize training. Meanwhile, label or feature skew, tiny per-client data shards, and domain or user shifts primarily amplify between-client heterogeneity. We therefore build Table~\ref{tab:src-benchmark-convergence} qualitatively through these factors when describing ``fast/slow/moderate'' and ``variance-prone'' behavior.

It is important to note that throughout our analysis, we analyze each trait in isolation by keeping the model, optimization, hyperparameters, training budget, and evaluation fixed. This clearly separates the effect of that trait on convergence, where any change in convergence reflects the influence of that single trait rather than other factors.

{\scriptsize
\setlength{\tabcolsep}{2pt}
\renewcommand{\arraystretch}{1.15}
\begin{longtable}{|
  >{\raggedright\arraybackslash}p{0.13\linewidth}|
  >{\raggedright\arraybackslash}p{0.27\linewidth}|
  >{\raggedright\arraybackslash}p{0.56\linewidth}|}
\caption{Standard Benchmark Sources: Convergence Profiles. X\%@rN is interpreted as the accuracy value reached at (@) round $N$. References with no $@rN$ mean they do not report the information of the global communication round or the converged round. \textit{Note: Model architectures and training configurations vary across the referenced studies.}}
\label{tab:src-benchmark-convergence}\\
\hline
\multicolumn{1}{|c|}{\textbf{Dataset}} &
\multicolumn{1}{c|}{\textbf{Convergence behavior}} &
\multicolumn{1}{c|}{\textbf{References / Evidence}} \\
\hline
\endfirsthead

\hline
\multicolumn{1}{|c|}{\textbf{Dataset}} &
\multicolumn{1}{c|}{\textbf{Convergence behavior}} &
\multicolumn{1}{c|}{\textbf{References / Evidence}} \\
\hline
\endhead

\endfoot

\hline
\endlastfoot

\textbf{MNIST} &
\textbf{Very fast / high.} Low-dim grayscale; clear margins; hit ceiling (asymptotic performance) fast. &
\textbf{FedICON}~\cite{tan2023heterogeneity}: 86.18--89.67\%@r100. \textbf{EvoFed}~\cite{rahimi2023evofed}: 97.62@r1000. \textbf{Every-Param}~\cite{zhou2024every}: 93.79--98.53\%@r100 (IID); 81.64--98.25\%@r100 (non-IID, max 5 labels/client); 71.64--95.85\%@r100 (non-IID, max 2 labels/client). \textbf{Stake-driven}~\cite{nguyen2024stake}: 87--88\%@r10. \textbf{pFedBreD}~\cite{shi2023prior}: 89.83--90.10\% (MCLR); 92.04--92.96\% (DNN). \par\smallskip
\textit{Summary:} median = 89.96\%, min--max = 71.64--98.53\%.
\\ \hline

\textbf{Fashion-MNIST} &
\textbf{Fast / moderate-high.} Same size as MNIST but finer textures $\rightarrow$ slightly slower/lower. &
\textbf{EvoFed}~\cite{rahimi2023evofed}: 84.72\%@r1000. \textbf{pFedBreD}~\cite{shi2023prior}: 98.44--98.51\% (MCLR); 98.73--98.98\% (DNN). \textbf{FeSEM-CAM}~\cite{ma2023structured}: 96.19--98.07\%@r100. \textbf{FedFed}~\cite{yang2023fedfed}: 94.01--94.31\%@r1000. \textbf{DSpodFL}~\cite{zehtabi2025decentralized}: 80\%@r3500 (IID); 70\%@r15000 (non-IID). \textbf{FedSAC}~\cite{wang2024fedsac}: 85.61--88.38\%@r100. \par\smallskip
\textit{Summary:} median = 90.58\%, min--max = 70.00--98.98\%.
\\ \hline

\textbf{EMNIST} &
\textbf{Moderate.} More classes increase confusion; low-dim $\rightarrow$ still stable. &
\textbf{pFedGate}~\cite{chen2023efficient}: 87.09--87.28\%@r400. \textbf{Flow}~\cite{panchal2024flow}: 90.88\%@r1500 (generalized); 94.18\%@r1500 (personalized). \textbf{Flash}~\cite{panchal2023flash}: 88.17\% (no drift); 89.18\%@r260 (incremental drift). \textbf{DoCoFL}~\cite{dorfman2023docofl}: 85.94--86.83\%@r10000. \textbf{PAdaMFed}~\cite{yan2025problem}: $\sim$90--92\% (IID); 87--90\% (non-IID). \par\smallskip
\textit{Summary:} median = 88.84\%, min--max = 85.94--94.18\%.
\\ \hline

\textbf{SVHN} &
\textbf{Moderate.} Complex background increase complexity. Converges better with augmentation. &
\textbf{FedSAC}~\cite{wang2024fedsac}: 70.51--81.95\%@r100. \textbf{FedICON}~\cite{tan2023heterogeneity}: 45.23\%@r100. \textbf{FedDPA}~\cite{yang2023dynamic}: 58.78--64.57\%. 
\par\smallskip
\textit{Summary:} median = 61.675\%, min--max = 45.23--81.95\%.
\\ \hline

\textbf{CIFAR-10} &
\textbf{Moderate.} Has more texture and colour vs MNIST. Converges with standard augmentation. &
\textbf{FedSAC}~\cite{wang2024fedsac}: 44.16--50.01\%@r100. \textbf{FedDisco}~\cite{ye2023feddisco}: 69.85--72.05\%@r100. \textbf{RECESS}~\cite{yan2023recess}: 60.85--64.32\% (under attack); 65.54\% (no attack). \textbf{pFedGate}~\cite{chen2023efficient}: 75.68--77.14\%@r400. \textbf{FLea}~\cite{xia2024flea}: 37.69--42.01\%@r100. \textbf{FedInit}~\cite{sun2023understanding}: 74.92--83.11\%@r500 (ResNet-18-GN); 79.73--88.47\%@r500 (VGG-11). \textbf{FedNLR}~\cite{wang2024fednlr}: 66.11--81.65\%@r500 (Dirichlet); 74.11--82.77\%@r500 (Shards). 
\par\smallskip
\textit{Summary:} median = 72.42\%, min--max = 37.69--88.47\%.
\\ \hline

\textbf{CIFAR-100} &
\textbf{Slow / low}. High sample complexity within the same class and less \#samples on each class. &
\textbf{FedNLR}~\cite{wang2024fednlr}: 44.67--51.77\%@r500 (Dirichlet); 47.93--52.46\%@r500 (Shards). \textbf{FedInit}~\cite{sun2023understanding}: 43.77--52.21\%@r500 (ResNet-18-GN); 50.27--58.84\%@r500 (VGG-11). \textbf{FedCOLLAB}~\cite{bao2023optimizing}: 40.94\%@r200. \textbf{Every-Param}~\cite{zhou2024every}: 37.40--67.81\%@r100 (IID); 24.98--66.74\%@r100 (non-IID, max 50 labels/client); 22.40--67.23\%@r100 (non-IID, max 20 labels/client). 
\par\smallskip
\textit{Summary:} median = 48.11\%, min--max = 22.40--67.81\%.
\\ \hline

\textbf{Tiny-ImageNet} &
\textbf{Slow / low}. Higher resolution with more classes demand stronger models or applying augmentation. &
\textbf{FedMRUR}~\cite{an2023federated}: 45.51--45.54\%@r1600 (Dirichlet); 45.42--45.71\%@r1600 (Pathological-n). \textbf{No One Idles}~\cite{zhang2023no}: SPFL: 60.21\%@r200 (IID), 48.31\% (non-IID); APFL: 60.18\%@r200(IID), 48.07\%@r200 (non-IID). \textbf{GuardHFL}~\cite{chen2023guardhfl}: 24.89--25.23\% (Q-priv); 25.82--28.46\% (Q-syn). \par\smallskip
\textit{Summary:} median = 46.82\%, min--max = 24.89--60.21\%.
\\ \hline

\textbf{FEMNIST (LEAF)} &
\textbf{Moderate--slow.} Tiny user shards and label skew. &
\textbf{pFedGate}~\cite{chen2023efficient}: 86.31--87.32\%@r400. \textbf{FedMR}~\cite{hu2024aggregation}: 83.76 $\pm$ 0.24\%@r1000 (CNN); 81.27 $\pm$ 0.31\%@r1000 (ResNet-20); 85.36 $\pm$ 0.21\%@r1000 (VGG-16). \textbf{RECESS}~\cite{yan2023recess}: 77.45--81.45\% (under attack); 82.30\% (no attack). \textbf{RACE}~\cite{liu2023one}: warm client: 65.93\%@r200 (global); 73.36\%@r200 (local). \textbf{FLuID} (Wang et al.): 80.1--81.1\%. \textbf{FedDPA}~\cite{yang2023dynamic}: 74.46--77.42\%. \par\smallskip
\textit{Summary:} median = 80.94\%, min--max = 65.93--87.32\%.
\\ \hline

\textbf{Shakespeare (LEAF)} &
\textbf{Slow / low.} Strong user non-IID slow stabilization compared to image classification. &
\textbf{FLuID}~\cite{wang2023fluid}: 41.7--43.6\%. \textbf{FedRoLa}~\cite{yan2024fedrola}: 47.42--48.06\%. \textbf{DoCoFL}~\cite{dorfman2023docofl}: 45.86--46.55\%@r10000. \textbf{Flow}~\cite{panchal2024flow}: 55.90\%@r1500 (generalized); 56.20\%@r1500 (personalized). \par\smallskip
\textit{Summary:} median = 47.74\%, min--max = 41.70--56.20\%.
\\ \hline

\textbf{Sentiment140 (LEAF)} &
\textbf{Moderate--Slow / variance-prone.} Sparse text and extreme user non-IID $\rightarrow$ high variance. &
\textbf{EPISODE++}~\cite{crawshaw2023federated}: 77.97\%@r2000. \textbf{pFedBreD}~\cite{shi2023prior}: 71.87--73.68\%. \par\smallskip
\textit{Summary:} median = 75.37\%, min--max = 71.87--77.97\%.
\\ \hline

\textbf{Synthetic (LEAF)} &
\textbf{Config-dependent.} Can be very slow under heavy skew or fast convergence under data homogeneity. &
\textbf{FedGMM}~\cite{wu2023personalized}: 72.02\%@r200 (Dir($\alpha$)=0.4, 300 clients, 200 rounds). \textbf{Flash}~\cite{panchal2023flash}: 93.92\%@r1000 (10/30 clients every round, 1000 rounds). \par\smallskip
\textit{Summary:} median = 82.97\%, min--max = 72.02--93.92\%.
\\ \hline

\textbf{Cora} &
\textbf{Fast / moderate} for GCN. Homophily aligns message passing with labels. &
\textbf{FedTGE}~\cite{wan2025energy}: 70.47--75.00\% (IID); 55.85--77.32 (non-IID). \textbf{HiFGL}~\cite{guo2024hifgl}: 85.55--86.42\%. \textbf{FedLoG}~\cite{kim2025subgraph}: 73.41--74.13\%@r100 (unseen node). \par\smallskip
\textit{Summary:} median = 73.25\%, min--max = 55.85--86.42\%.
\\ \hline

\textbf{CiteSeer} &
\textbf{Moderate.} Higher dim + lower homophily than Cora $\rightarrow$ make it slower/harder. &
\textbf{HiFGL}~\cite{guo2024hifgl}: 77.24--77.91\%. \textbf{FedLoG}~\cite{kim2025subgraph}: 66.34--76.45\%@r100 (unseen node). \par\smallskip
\textit{Summary:} median = 74.485\%, min--max = 66.34--77.91\%.
\\ \hline

\textbf{PubMed} &
\textbf{Moderate.} Larger graph. Homophily helps, but the imbalance slow training a bit. &
\textbf{FedTGE}~\cite{wan2025energy}: 57.19--85.79\% (IID); 67.22--86.79\% (non-IID). \textbf{HiFGL}~\cite{guo2024hifgl}: 85.04--86.26\%. \textbf{FedLoG}~\cite{kim2025subgraph}: 86.22--86.27\%@r100 (unseen node). \par\smallskip
\textit{Summary:} median = 81.3275\%, min--max = 57.19--86.79\%.
\\ \hline

\end{longtable}
}

{\scriptsize
\setlength{\tabcolsep}{2pt}
\renewcommand{\arraystretch}{1.2}
\begin{table}[htbp!]
\centering
\caption{Summary of data traits and its convergence impact}
\label{tab:traits-convergence}
\scriptsize
\setlength{\tabcolsep}{2pt}
\renewcommand{\arraystretch}{1.05}
\begin{adjustbox}{max size={\linewidth}{\textheight}}
\begin{tabular}{|
  >{\raggedright\arraybackslash}p{0.15\textwidth}|
  >{\raggedright\arraybackslash}p{0.25\textwidth}|
  >{\raggedright\arraybackslash}p{0.47\textwidth}|
  >{\centering\arraybackslash}p{0.08\textwidth}|}
\hline
\multicolumn{1}{|c|}{\textbf{Trait}} &
\multicolumn{1}{c|}{\textbf{Meaning}} &
\multicolumn{1}{c|}{\textbf{Convergence effect}} &
\multicolumn{1}{c|}{\textbf{Impact}}\\
\hline
\textbf{Provenance (Cross-X)} & Cross-silo (few, reliable) vs cross-device (many, intermittent). & Cross-device: more heterogeneity, higher dropout, lower per-round participation $\rightarrow$ slower and less stable. Cross-silo $\rightarrow$ steadier and faster. & Strong \\
\hline
\textbf{\#Dimension} & Input dimensionality (resolution or feature size). & Higher dimension $\rightarrow$ lower accuracy and slower convergence. & Medium \\
\hline
\textbf{\#Samples} & Number of samples. & More samples helps only when it increases data volume or label coverage among participating clients; otherwise has little impact. & Light \\
\hline
\textbf{\#Classes} & Number of target categories. & More classes increase confusion $\rightarrow$ lower accuracy and slower convergence. & Medium \\
\hline
\textbf{Native label skew} & Imbalance of labels/distributions across clients. & More skew $\rightarrow$ lower accuracy (at finite rounds) and slower convergence. & Medium \\
\hline
\textbf{ICV} & Diversity within each class. & Higher ICV $\rightarrow$ different features within same class $\rightarrow$ lower accuracy and slower convergence. & Strong \\
\hline
\textbf{ICS} & How far apart classes are (margin). & Larger ICS $\rightarrow$ clearer features between classes $\rightarrow$ easier to separate $\rightarrow$ higher accuracy and faster convergence. & Strong \\
\hline
\textbf{Graph homophily} & Likelihood that connected nodes share labels. & Higher homophily $\rightarrow$ faster, more stable convergence with little accuracy gain. & Medium \\
\hline
\end{tabular}
\end{adjustbox}
\end{table}
}

\subsubsection{Provenance} 
Evidence from systems studies and our compiled benchmarks converges on the same picture: the origin and operating context of the data (cross‑silo vs cross‑device), primarily shifts convergence speed and stability, with secondary effects on the final ceiling that become pronounced under finite communication budgets. Comparative analyses report that cross‑silo FL usually involves few, reliable clients (2–100), whereas cross‑device FL may span very large populations (up to $10^{10}$ devices) with slower links, intermittent availability, and higher dropout, yielding slower and less stable convergence~\cite{kairouz2021advances}. Google's system‑scale observations mirror this: cross‑device settings converge more slowly and with higher variance because statistical heterogeneity and participation rate are both larger~\cite{bonawitz2019towards}. This systems‑level diagnosis is consistent with the algorithmic literature showing FedAvg slowdowns under non‑IID sampling and proposing correctives such as SCAFFOLD~\cite{karimireddy2020scaffold}, FedProx~\cite{li2020federated}, and FedNova~\cite{wang2020tackling} to counter gradient bias and client drift. The empirical patterns in Table~\ref{tab:src-benchmark-characteristics} and Table~\ref{tab:src-benchmark-convergence} reflect these claims across modalities. In cross‑device image and text sources, native user partitioning amplifies skew and variance: FEMNIST needs $\sim$ 400–1000 rounds to stabilize and plateaus lower (median 80.94\%) relative to EMNIST (median 88.84\%) despite comparable total samples; Shakespeare remains in the mid‑40s to mid‑50s even after 1500–10,000 rounds, while Sentiment140 reaches mid‑70s but typically requires thousands of rounds when client sampling is thin. In contrast, cross‑silo graph benchmarks exhibit steadier, faster progress at similar round budgets because client populations are small and reliable and per‑round participation is high: Cora often stabilizes by $\sim$100 rounds with medians in the low‑to‑mid‑70s; PubMed attains mid‑80s (and $\approx$86\% for unseen nodes by $r=100$ when homophily is preserved. This aligns with reports from medical multi‑center (cross‑silo) studies, where training across tens of hospitals achieves stable convergence and strong final accuracy; the principal challenge is institutional heterogeneity, but with near‑full participation, the optimization remains tractable~\cite{darzidehkalani2022federated, linardos2022federated}. Finally, for ``generic'' image sources (MNIST, CIFAR‑10/100, Tiny‑ImageNet) that are often synthetically partitioned, provenance matters less than inter‑class separation and intra‑class variability: at fixed recipes, MNIST rises to high‑90s within ($\mathrm{O}(10^2)$) rounds because margins are large, whereas Tiny‑ImageNet (fine‑grained, higher resolution) hovers in the mid‑40s even after hundreds–1600 rounds—illustrating that geometry--not just provenance--controls the ceiling.

Overall, in this comparison, cross‑silo FL is typically faster and more stable because of fewer, more reliable clients and higher per-round participation, while cross‑device FL is slower and less stable due to greater statistical heterogeneity and dropout rate.

\subsubsection{Feature dimension}
Dimensionality interacts with class geometry and capacity to influence both speed and ceiling, but its effect is secondary to separation and variability. Among images, moving from $1\times28\times28$ grayscale digits (MNIST, median 89.96\%) to $3\times32\times32$ color natural images (CIFAR-10, median 72.42\%) slows learning and lowers the plateau at comparable rounds; pushing to $3\times64\times64$ (Tiny‑ImageNet, median 46.82\%) accentuates this drop unless capacity and augmentation scale accordingly. The same dimensional rise in graphs does not preclude strong outcomes when structure helps: PubMed (500-d TF-IDF) attains around 86\% by $r = 100$ in several reports, whereas Cora (1433‑d BoW) ranges widely depending on partitioning and method. These contrasts indicate that higher dimension raises sample complexity and gradient variance per round but does not deterministically cap accuracy; the decisive factor is whether increased dimension comes with tighter margins or, as in Tiny-ImageNet, substantially higher within‑class scatter.

From experimental views, FedScale\cite{lai2022fedscale} measures on-device training latency and shows that for the same model and same mini-batch size, larger training images take longer execution per step. In particular, ImageNet mini-batches run slower than CIFAR-10 on the same devices, where ImageNet images normally have a much larger size. Besides, in a cross-site chest X-ray study\cite{slazyk2022cxr}, models trained on segmented lung regions performed worse than the same architectures trained on full images. In here, accuracy may improve only if the extra pixels add task-relevant signal; otherwise, it just adds more time. So, image resolution is a first-order system factor, where higher resolution leads to higher FLOPs/batch on clients, resulting in convergence with longer rounds and slower wall-clock convergence.

\subsubsection{Number of samples and number of classes} 
Sample count is a weak factor once class geometry (e.g., IVS, ICS, per-client coverage and overlap) and skew are fixed. EMNIST and FEMNIST both have $\approx$800k images, yet the latter converges slower and lower at practical round budgets because writer‑sharding amplifies drift (80.94\% median vs 88.84\% for EMNIST, with more rounds needed to stabilize). Conversely, Tiny‑ImageNet has more total samples (120k) than CIFAR‑10 (60k) but achieves a lower median (46.82\% vs 72.42\%) due to finer granularity and higher resolution that reduce margins and raise within‑class spread. Graph benchmarks reinforce the point: PubMed's larger node set (19,717) does not impede mid‑80s accuracy when structural alignment is helpful, while Cora, though smaller, can underperform under non‑IID partitions. In short, more data helps if and only if it increases margin or reduces variance at the client level; otherwise, it mainly raises per‑round compute without accelerating convergence.

Increasing the number of classes predominantly lowers inter‑class separation and elevates confusion, which slows convergence and depresses the ceiling at fixed model capacity. The cleanest comparison is CIFAR‑10 (10 classes; median 72.42\%) to CIFAR‑100 (100 classes; median 48.11\%) under similar resolutions and rounds, with Tiny‑ImageNet (200 classes) pushing the trend further (median 46.82\% and often hundreds to 1600 rounds). The text benchmarks echo this: Sentiment140's binary polarity exhibits a markedly higher median (75.37\%) than Shakespeare's next‑character task, where effective ``class'' alternatives per step are large and signals sparse, yielding 41.7–56.2\% even with long training. These observations are consistent with the view that the number of classes affects optimization primarily by shrinking margins and increasing the sample complexity needed per decision boundary.

\subsubsection{Label skew}

Native label skew—prominent in cross-device sources-principally slows convergence by misaligning client gradients and inflating variance; its effect on final accuracy becomes visible under finite round budgets. FEMNIST, sharded per writer, requires $\approx$400–1000 rounds to stabilize and still sits $\approx$8 points below EMNIST's median despite comparable total size, whereas EMNIST reaches low-90s with sufficient rounds or personalization. In text, Shakespeare's combination of extreme user skew and weak margins leads to very slow, low plateaus even with 1,500-10,000 rounds, while Sentiment140, though also skewed, fares better because the binary signal offsets drift. The Synthetic LEAF studies make this lever explicit: tightening the Dirichlet $\alpha$ (stronger skew) slows and lowers outcomes at fixed rounds, while increasing the clients‑per‑round or $\alpha$ relaxes the penalty. 

\subsubsection{ICV and ICS}
Higher within‑class scatter inflates the local Lipschitz landscape and client‑to‑client gradient disagreement, which both delays stabilization and reduces the plateau unless capacity scales. The empirical gradient is clear across images: SVHN's cluttered backgrounds yield a median of 61.68\% with a wide spread by $r\approx100$ (45–82\%), while CIFAR‑10's natural textures sit higher at 72.42\% and MNIST's low‑variability digits much higher still (89.96\%). Moving to CIFAR‑100 and Tiny‑ImageNet raises within‑class variability through fine‑grained distinctions, dropping medians to 48.11\% and 46.82\% despite far more rounds. On FEMNIST, writer‑specific glyph styles increase within‑class dispersion relative to EMNIST, requiring hundreds more rounds to reach a lower plateau. These consistent declines with increasing variability point to ICV as a primary brake on both learning speed and attainable accuracy. Given that, reducing ICV is always the first aim of any approaches when working with long-tail distribution~\cite{yan2022age}.

Larger margins accelerate progress and raise asymptotic accuracy at fixed capacity, which is visible whenever we can hold other traits roughly constant. MNIST vs Fashion‑MNIST isolates this effect: both are $28\times28$, ten‑class problems, yet the finer textures of Fashion‑MNIST reduce effective separation; outcomes span 70–99\% and stabilization often needs more rounds than MNIST's $\approx$100‑round fast rise to high‑90s. In text, Sentiment140's binary polarity provides stronger separation than next‑character modeling, yielding a median of 75.37\% versus Shakespeare's 47.74\% despite comparable cross‑device skew. In images, the pronounced decline from CIFAR‑10 to CIFAR‑100 at the same resolution and similar rounds reflects a systematic compression of inter‑class distances as granularity rises. Where methods enlarge margins (e.g., stronger backbones or tailored augmentation), the same datasets move from mid‑70s to upper‑80s by $r\approx500$, underscoring that separation chiefly determines both the trajectory and the ceiling.

\subsubsection{Graph homophily}
On graphs, homophily is the structural analogue of ICS: it aligns message passing with label smoothness, improving gradient alignment and speeding convergence. Cora's high homophily allows stabilization by 100 rounds with typical accuracies in the low‑to‑mid‑70s and frequent reports above 80\% under favorable partitions; CiteSeer exhibits lower homophily and correspondingly slower, more method‑sensitive progress at similar rounds and medians in the mid‑70s. PubMed, with moderate‑to‑high homophily and class imbalance, still reaches mid‑80s and attains 86\% by $r=100$, indicating that structural alignment can offset scale. 
\vspace{0.15cm}

\noindent\textbf{Methodological debates and mixed findings:} Throughout the analysis, we observe substantial cross-paper dispersion in convergence behavior, even on the same benchmark with similar trait levels or reported rounds. This can lead to various contradictory conclusions about whether a trait mainly affects convergence speed, stability, or the final ceiling. For example, SVHN at round 100 ranges from 45.23\% (FedICON \cite{tan2023heterogeneity}) to 70.51--81.95\% (FedSAC~\cite{wang2024fedsac}), or CIFAR-10 spans 37.69--88.47\% across references. Graph benchmarks also show large spreads between PubMed, ranging from 57.19 to 86.79\%, and Cora from 55.85 to 86.42\%. These discrepancies are not necessarily true conflicts. Instead, they often arise from differences in experimental settings and in how traits are instantiated.

Given the note about setting differences in Table 2, we briefly mention the three main factors (may/may not be data-related) varying across studies that can shift the trait effects to the final ceiling and convergence behavior:
\begin{itemize}
    \item \textit{Skewness parameter:} Skewness can arise at different levels. Yet, scoping it into the data level only, a clear observation is that severe non-IID parameterization not only reduces the final ceiling, but also slows down convergence, when all other settings are fixed. Under a defined round budget, this skewness can substantially change the measured accuracy. For example, on MNIST, Every-Param~\cite{zhou2024every} reports 93.79–98.53\% at round 100 under IID, but only 71.64\% at the same round when each client is restricted to at most two labels (non-IID). Besides, even with a much larger training budget, non-IID data may not catch up to IID performance. On Fashion-MNIST, DSpodFL~\cite{zehtabi2025decentralized} reaches 80\% at round 3500 under IID, whereas only 70\% after taking 4$\times$ more rounds under non-IID.

    \item \textit{Budget and additional constraints:} The differences in convergence can sometimes be driven by threat or privacy settings rather than intrinsic data difficulty. As can be seen, the reported final ceilings differ substantially when each work has its own training budget and environmental constraints. Some works report results at fixed rounds, others at longer horizons (e.g., thousands of rounds), and some operate under constraints such as attacks or privacy mechanisms. For example, in CIFAR-10 and FEMNIST, RECESS~\cite{yan2023recess} reports degraded outcomes under attack relative to the no-attack case, while in Tiny-ImageNet, GuardHFL~\cite{chen2023guardhfl} under quantized privacy settings yields markedly lower accuracies than unconstrained baselines in the same benchmark family. 
    
    \item \textit{Model choices:} This is particularly evident when the same dataset shows large performance spreads at similar communication rounds, or when changing the model backbone leads to markedly different outcomes. Such variability is reported in prior work that evaluates the same setting with different architectures. For example, on CIFAR-10, FedInit~\cite{sun2023understanding} reports substantially different performance ranges for ResNet-18-GN and VGG-11 at 500 rounds. This aligns with our geometric view, where traits such as ICV/ICS capture intrinsic aspects of task difficulty, but the effective margin and within-class scatter can be partially modified and shaped by pre-processing and model optimization choices. Consequently, both the convergence trajectory and the performance ceiling can shift even under the same data distribution and a fixed training budget.
\end{itemize}

\subsection{Traits relation and impact ranking}
\begin{figure}[h!]
\centering
\includegraphics[width=.9\textwidth]{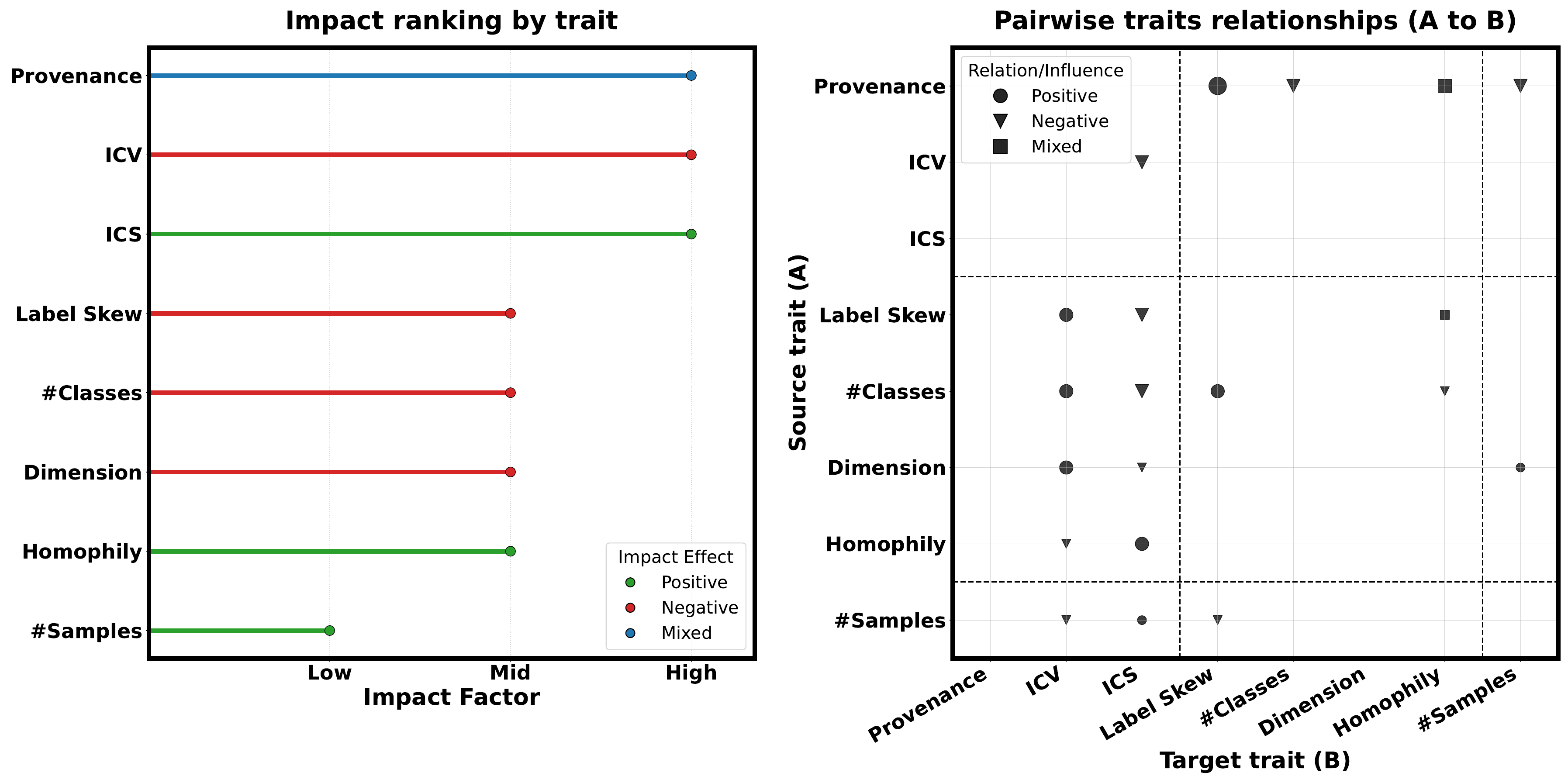}
\caption{Data traits: impact on convergence behaviors (left), and pairwise relations (right).}
\label{fig:src-plots}
\end{figure}

In a dataset, these traits are somewhat interdependent and not isolated. Figure \ref{fig:src-plots} (left) arranges the traits by their qualitative impact on convergence into three tiers: high, mid, and low. It uses color to indicate each trait’s overall tendency with respect to convergence outcomes, where green denotes a generally beneficial tendency, red a detrimental tendency, and blue refers to a mixed effect. The subplot on the right of Fig. \ref{fig:src-plots} summarizes directed pairwise relations among traits. An entry (displayed by a marker) at row A, column B indicates that increasing source trait A tends to aid/support, eliminate, or contextually alter target trait B, while the marker size is proportional to effect strength: light, medium, strong. This illustration exposes the pathways by which traits influence one another, complementing the left's ranking of their overall impact on convergence.

At the top of the high impact factor are provenance, ICV, and ICS. Together, they determine how aligned the gradients are. When provenance shifts from cross-silo to cross-device, clients typically hold fewer examples and a narrower set of labels. The immediate consequence is greater label skew and smaller per-client class coverage, which in turn amplifies client drift. Geometry (ICV, ICS) then mediates the remaining effects. Increases in dimensionality (a 256x256 compared to a 28x28 image) give more places for the same-class examples to vary, which easily inflates ICV. At the same time, classes can look closer together because many classes must share a crowded feature space, which reduces effective ICS. 

The number of classes drives a similar pattern. As classes proliferate under a fixed data budget, the long tail grows, measured skew rises, within-class structure becomes more multi-modal (ICV increases), and pairwise margins compress (ICS decreases). In graph domains, homophily acts as a geometric amplifier. Stronger homophily typically sharpens class separation after message passing, which raises effective ICS, and it can damp local within-class noise, which lowers ICV. 

Data volume (\#samples) primarily improves estimation rather than changing the intrinsic difficulty. With more samples, centroids and covariances are estimated more reliably, which tightens apparent ICV and modestly enlarges effective ICS. Also, small differences between classes stop looking like big imbalances when we have a larger amount of data. As observed, we can affirm that the augmented data helps, but it can rarely overturn the dominant influences set by provenance and geometry.

Overall, the net effect of a single trait can change depending on the level of the others. Yet, when capacity and hyperparameters are controlled, final accuracy and convergence speed are governed first by ICS (higher is better) and ICV (lower is better); native label skew and graph homophily then modulate speed by changing gradient alignment across clients and partitions, with accuracy penalties emerging at realistic (finite) round budgets. Dimensionality, sample size, class count, and task form matter insofar as they alter separation and variability or interact with them. For example, higher resolution and more classes often compress margins and inflate scatter. 
The empirical contrasts: MNIST versus Fashion‑MNIST and SVHN, CIFAR‑10 versus CIFAR‑100 and Tiny‑ImageNet, EMNIST versus FEMNIST, Shakespeare versus Sentiment140, Cora/CiteSeer/PubMed consistently support this hierarchy (\textit{geometry first, heterogeneity second, scale last}).


\section{Data splitting}
\label{sec:spliting}

Many prior works have focused on data splitting strategies for centralized ML models, specifically on how to effectively divide datasets into training and testing sets~\cite{medar2017impact, vrigazova2021proportion, racz2021effect, tan2021critical, kahloot2021algorithmic, salazar2022fair, muraina2022ideal, joseph2022split}. These studies have successfully explored the impact of different splitting proportions and techniques on model performance in traditional setups. Yet, there is a noticeable gap between these and research on effective data splitting across clients in the FL context, where data is distributed across decentralized devices with various complex challenges in disparity and privacy. This lack of focus on client-based data splitting in FL highlights a need for further development of strategies that address the distinctive challenges in federated environments. 

In this section, we present the factors that make it particularly difficult to design distribution methods that accurately reflect real-world data distributions (\textit{RQ2}) and introduce current approaches for simulating as best as possible such distributions in FL (\textit{RQ3}). We also present recent approaches that try to mitigate the impact of \textit{non-IID} data on the model convergence.

\subsection{The challenges of real-world data distribution}

In the FL context, real-world data is often non-IID across clients, meaning that different clients may have data with varying distributions, feature spaces, or data volumes. However, most prior research relies only on simulation using standardized benchmark datasets like \href{http://yann.lecun.com/exdb/mnist/}{MNIST} and \href{https://www.cs.toronto.edu/~kriz/cifar.html}{CIFAR}, equally splits them then tests or compares model performance. This improper splitting may not fully capture the complexities of diversity seen in practice. However, if a simulation tries to simulate the disparity across clients with too much variation, it may cause difficulty in comparing models in experiments and doing it under controlled conditions. Consequently, the main challenge in FL-related research is how to split data in a way that mimics and reflects the heterogeneity of real-world FL. Supporting this statement, McMahan et al.~\cite{mcmahan2017communication} and Zhao et al.\cite{zhao2018federated} have shown in their works that unbalanced or highly skewed non-IID data can negatively degrade model performance and cause more struggles if not carefully handled.

To break the term "heterogeneity" into smaller problems, we can discuss the difficulty of disparity from several views: quantity, type of data, and other subjective factors across clients. About data quantity, some clients might have significantly more data than others. This disparity can lead to uneven contributions during training, where models may become biased toward clients with larger datasets, and clients with less data are likely to have slower convergence~\cite{li2020federatedsurvey}. Simulating this quantity imbalance effectively, without skewing the results or favoring certain clients, remains a challenge in finding the optimal splitting method.  

Clients may also differ in terms of the type of data they handle. In real-world FL systems, different clients may work with different domains or modalities of data. For example, one client may collect text data, while another might gather images or sensor data. In simulations, finding a way to partition datasets to reflect the cross-domain and cross-modal data without oversimplifying the problem (e.g., using assumptions) is difficult. 
Besides, this feature-partitioned setting, often referred to as vertical FL (VFL), poses additional challenges that can directly affect convergence. Unlike horizontal FL, where clients usually share the same feature space but have different samples, VFL involves parties that hold different feature subsets for the same or partially overlapping entities~\cite{khan2025vertical}. Therefore, the learning process depends on reliable entity alignment. If the overlap among parties is limited, the effective training set becomes smaller and may no longer represent the original population well. If the alignment is imperfect, features from one party may be associated with incorrect labels or inconsistent feature blocks from another party, which introduces noisy gradients and can slow convergence or guide the model toward a biased solution~\cite{liu2024vertical}. Feature heterogeneity further complicates this process because different parties may hold features with different scales, sparsity levels, missing-value patterns, modalities, and categorical encoding. This can create imbalanced gradient contributions across parties, make the optimization problem poorly conditioned, and require additional normalization, calibration, or party-specific weighting to stabilize training. Moreover, VFL is commonly implemented with privacy-preserving protocols, homomorphic encryption, or secure multi-party computation to protect raw data and intermediate information, even though they can significantly increase communication and computation costs, introduce synchronization delays, and straggler effects~\cite{liu2024vertical, ye2025vertical}.
Due to those limitations, most existing benchmark datasets typically do not account for multimodal data or Vertical FL, making it harder to simulate and evaluate models that are expected to handle diverse data types in real federated deployments. This statement aligns with the observation made in FedMultimodal by Feng et al.~\cite{feng2023fedmultimodal}, where the authors discuss that most existing studies conduct experiments under a unimodal setting.

Finally, one external factor creating this challenge is dynamic client participation, where not all clients participate consistently. Some may drop out or rejoin, or they may only intermittently contribute. Simulating this dynamic participation is difficult, especially when it comes to deciding how to split or re-distribute data when clients leave or join the training process~\cite{zhang2022federatedsurvey, xu2023asynchronous}.

Each of these challenges adds layers of complexity to data splitting in FL simulations, beyond simply replicating real-world data heterogeneity.

\subsection{Data splitting in federated settings}

In the previous section, we discussed the challenges surrounding data splitting in FL and how the heterogeneity of client data complicates the simulation of real-world scenarios. Building on that, this section will explore how data is typically split for FL settings.

\subsubsection{Dirichlet-distribution-based splitting} Many recent FL studies~\cite{liu2024badsampler, wang2024fednlr, wang2024fedsac, shejwalkar2022back, morafah2023flis, cai2023fedce} evaluated the reliability of their works on data distributed according to a Dirichlet distribution~\cite{connor1969concepts, ng2011dirichlet, ongaro2013generalization}, as it has been shown to closely resemble real-world scenarios and make the benchmark datasets more realistic for federated settings. To be more specific, datasets like MNIST or CIFAR instead of equally distributed to all clients with the same number of samples or classes, will be skewly distributed based on the $\alpha$ value. This factor controls the concentration of data across clients, in which a smaller $\alpha$ leads to a more skewed distribution, and a larger value of $\alpha$ results in a more uniform distribution across clients. 

\begin{figure}[h!]
\centering
\includegraphics[width=.65\textwidth]{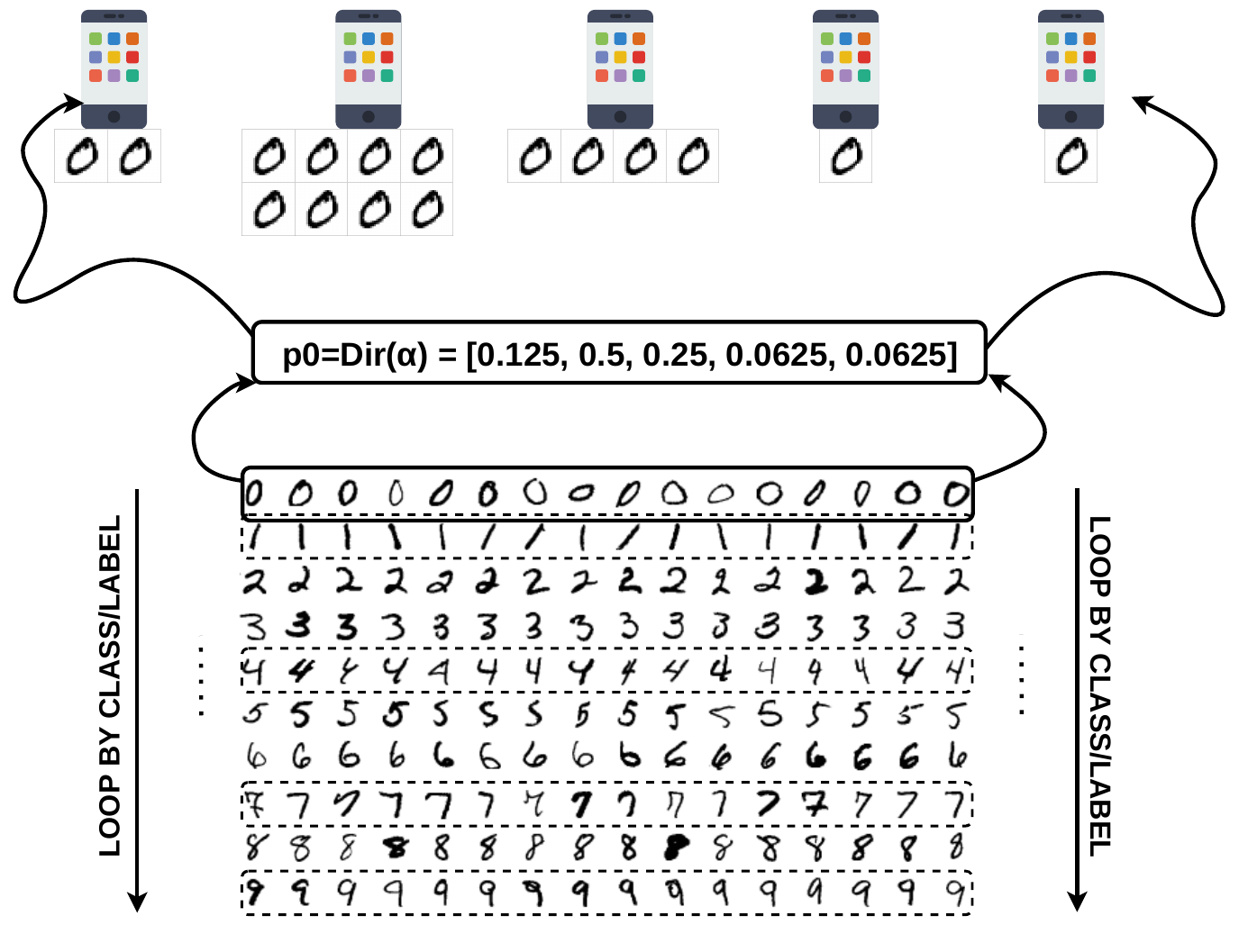}
\caption{Dirichlet-distribution-based splitting}
\label{fig:split-dirichlet}
\end{figure}

Let's assume there are $N$ clients in the system and $C$ classes in the dataset. For each class $c \in C$, we randomly draw or sample the probability distribution of that class across clients. For example,  in the case N=5, this sampled vector can be [0.15,  0.52, 0.1, 0.2, 0.03], meaning that 15\% of the data goes to client 1, 52\% goes to client 2, and so on. Accordingly, client 2 will have the largest number of samples of class $c$ while client 5 faces scarcity with only 3\% of the data of this class distributed to it. This process is repeated for all classes in the dataset to ensure that each client ends up with a non-IID mix of different classes, based on the $\alpha$ factor of the Dirichlet distribution. Finally, as previously mentioned, the degree of skew varies with different $\alpha$ values, leading to differences in the final number of samples assigned to each client. This can effectively simulate real-world federated scenarios. See Fig. \ref{fig:split-dirichlet} for an illustration of how data is split according to the Dirichlet distribution. 

In practice, fixed-$\alpha$ settings often yield faster and smoother convergence than real deployments because they primarily perturb label proportions while leaving other factors (like client-size long tails and participation dynamics) under-modeled. This can overestimate accuracy and stability. For more faithful stress tests, vary $\alpha$ and impose a realistic client-size distribution, minimum samples per client/class, and partial participation. Always report $\alpha$, seeds, per-client class coverage, and the client-size histogram.

\subsubsection{Natural splitting}\label{sec:natural-splitting}
\begin{figure}[h!]
\centering
\includegraphics[width=.67\textwidth]{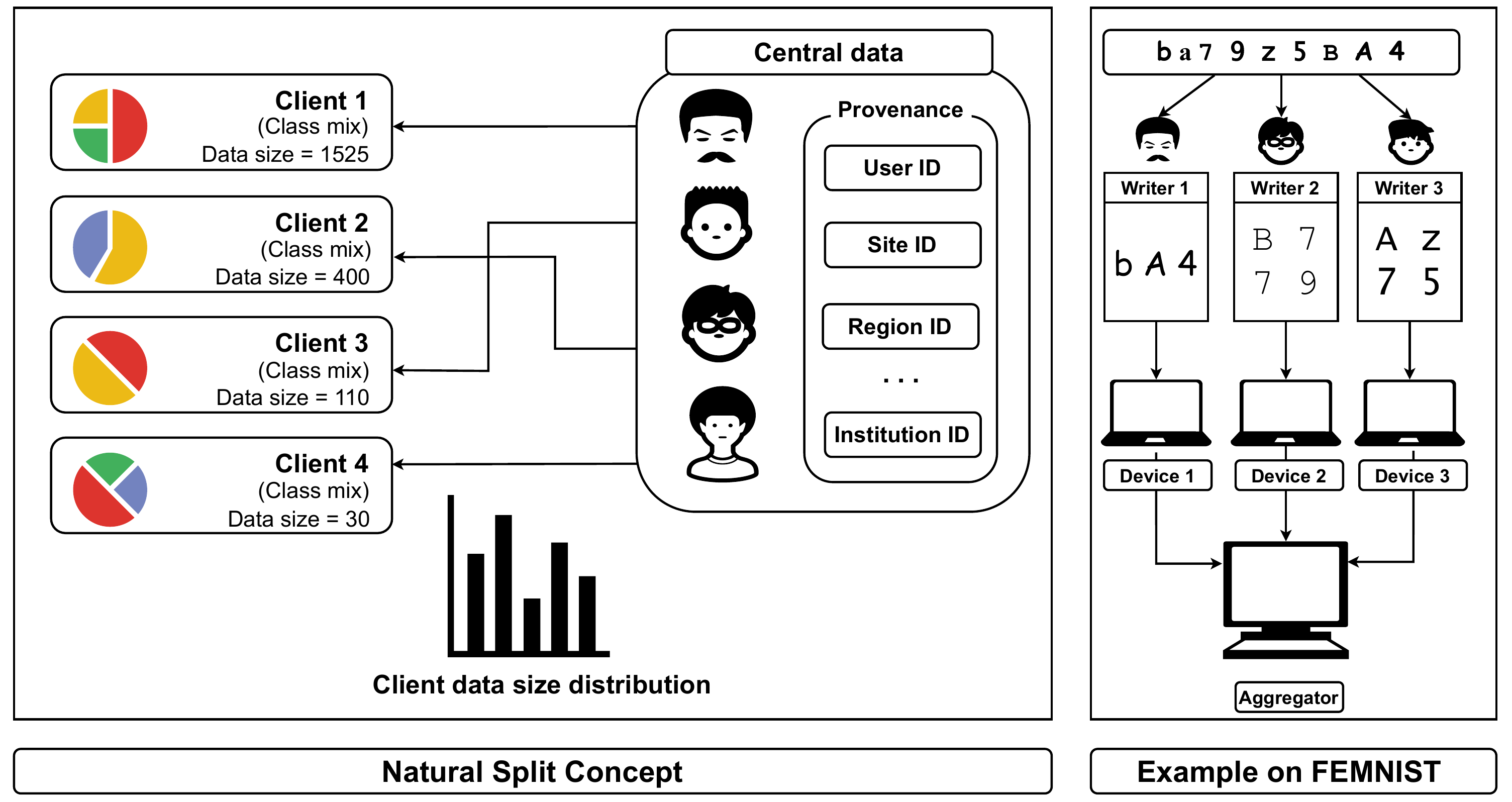}
\caption{Natural splitting}
\label{fig:split-natural}
\end{figure}

A \emph{natural} split partitions data by its real-world provenance (e.g., user or device identifier, site or hospital, sensor or producer, or geographic region) and preserves the original client boundaries without re-mixing examples to meet a target class balance (see Fig. \ref{fig:split-natural}). In contrast to synthetic procedures such as Dirichlet-$\alpha$ label allocation, natural splits inherit the coupled sources of heterogeneity that occur in practice, including multimodal client populations, long-tailed client sizes, correlated covariates and labels, missing classes, and acquisition or workflow artifacts. As a result, evaluation under natural splits is typically more faithful to deployment conditions, albeit less controllable and sometimes harder to compare across papers than a fixed-$\alpha$ setting. A notable example in experimental datasets is \href{https://leaf.cmu.edu/}{FEMNIST} (set of letters and digits), naturally non-IID with data explicitly grouped by the writer ID. Since each writer has a unique writing style, this partitioning method ensures the distinctiveness of partitioned data for each client and aligns more closely with realistic FL scenarios. In a similar vein, LEAF adopts this partitioning scheme to split Sentiment140 and Reddit by user or user ID, while CelebA is partitioned by identity, rather than following the conventional train/validation/test split used in the original paper~\cite{liu2015faceattributes}.

Several benchmark efforts explicitly advocate for and implement natural splitting toward the application side. FedScale constructs client-aware partitions from raw metadata and releases workloads where each client corresponds to a real data owner, together with trace-driven participation patterns; the authors caution that purely synthetic partitions can mask systems and statistical challenges that appear under genuine client boundaries~\cite{lai2022fedscale}. FLamby provides multi-site healthcare datasets with site-level partitions and per-client folds, arguing that natural site splits better capture domain-specific heterogeneity than synthetic label shuffles and thus yield safer cross-silo benchmarks~\cite{ogier2022flamby}. When only centralized proxy data are available, a proposed approach is learning mixtures of Dirichlet-multinomial models from distributions observed in live federated deployments and then sampling simulated client histograms to partition the proxy data accordingly; this learned-from-nature simulation improves fidelity over IID or naive Dirichlet baselines and is compatible with privacy-preserving aggregation~\cite{scott2024improved}.

Based on the studied works, natural client partitions generally converge slowly and destabilize convergence relative to IID. We recommend prioritizing natural client boundaries over synthetic Dirichlet-distribution-based splitting whenever possible and reporting quantitative indicators of heterogeneity to make results interpretable and comparable. For completeness and ablation, including a synthetic Dirichlet baseline with clearly reported $\alpha$ remains useful. However, relying solely on a fixed-$\alpha$ partition risks underestimating optimization difficulty and overestimating accuracy under non-IID conditions. This hybrid protocol, with natural split as the primary setting and synthetic partitions for controlled analysis, balances external validity with diagnostic clarity.

\subsubsection{Cluster-based splitting} 
\begin{figure}[h!]
\centering
\includegraphics[width=.75\textwidth]{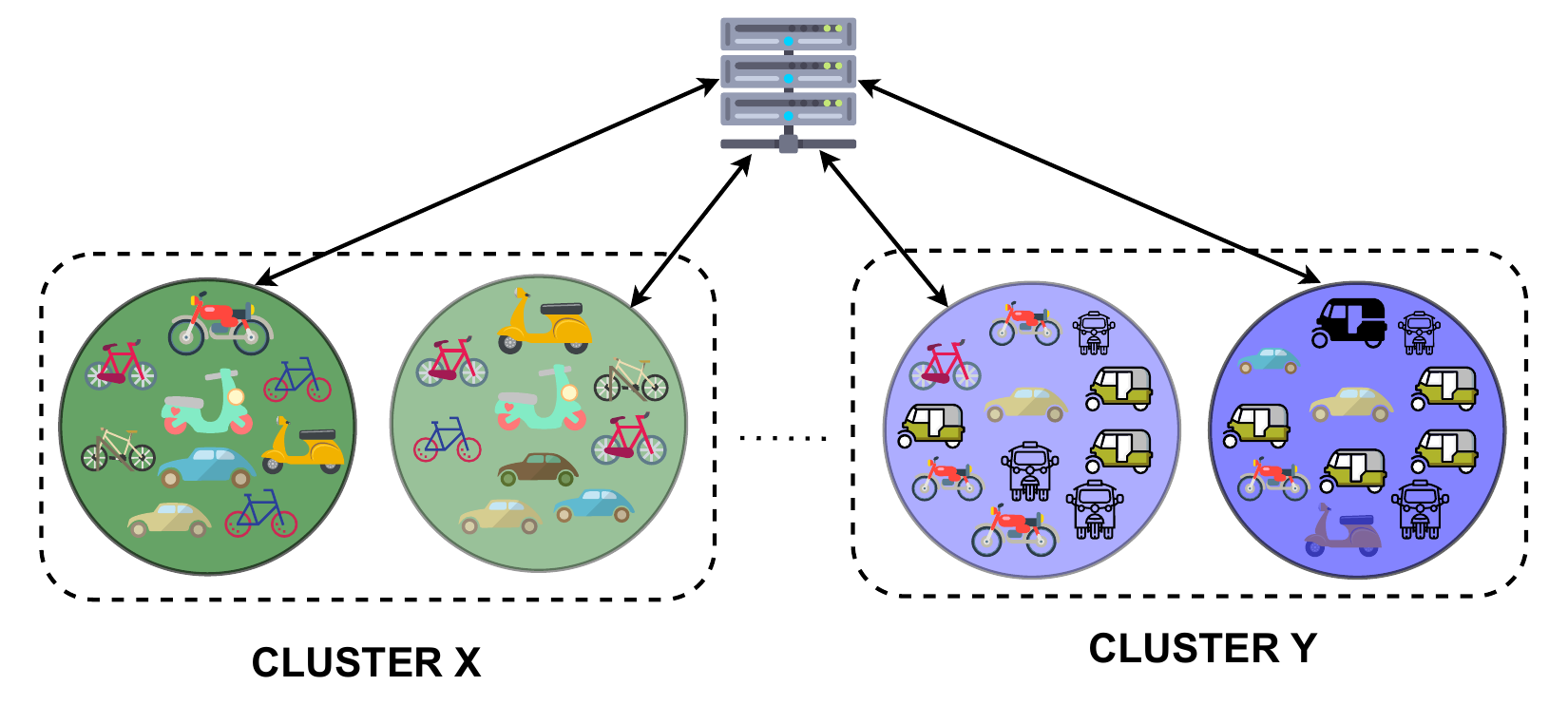}
\caption{Cluster-based splitting}
\label{fig:split-cluster}
\end{figure}

More manually, data is split using other criteria. For instance, it can be based on geographical or regional clusters, where clients within the same cluster may have similar data distributions (see Fig. \ref{fig:split-cluster}). In practice, traffic flow prediction across cities is one potential application of this splitting method. Each city would train its model using local data collected from roadside or intersection cameras. This may lead to data heterogeneity, as one city might have more images of tuk-tuks (one common vehicle in regions like Thailand, Indonesia, India, and Bangladesh), while another has more car images. If the model is trained to detect vehicles and predict traffic flow based on those detections, FL could enable knowledge sharing or collaboration among cities (acting as clients in the FL system), improving detection performance in any vehicle type. Despite many prior works considering the geography or similarity relation in their FL research~\cite{park2023fedgeo, yin2020fedloc, tan2024bridging, li2021federated, duan2021flexible, long2023multi, morafah2023flis}, yet, only when the proposed algorithm is tested on data split according to a specific real-world problem and its distribution, the evaluation results can be considered relevant and applicable in practice later. 

Under a single global model, optimization may oscillate or converge slowly when clusters are far apart, and the global optimum can underfit minority clusters. Clustered FL methods often converge faster within each mode. To keep evaluations sound, define clusters without label leakage, report the clustering signal and stability, and quantify inter- vs.\ intra-client divergence.

\subsubsection{Temporal-based splitting} 
\begin{figure}[h!]
\centering
\includegraphics[width=.65\textwidth]{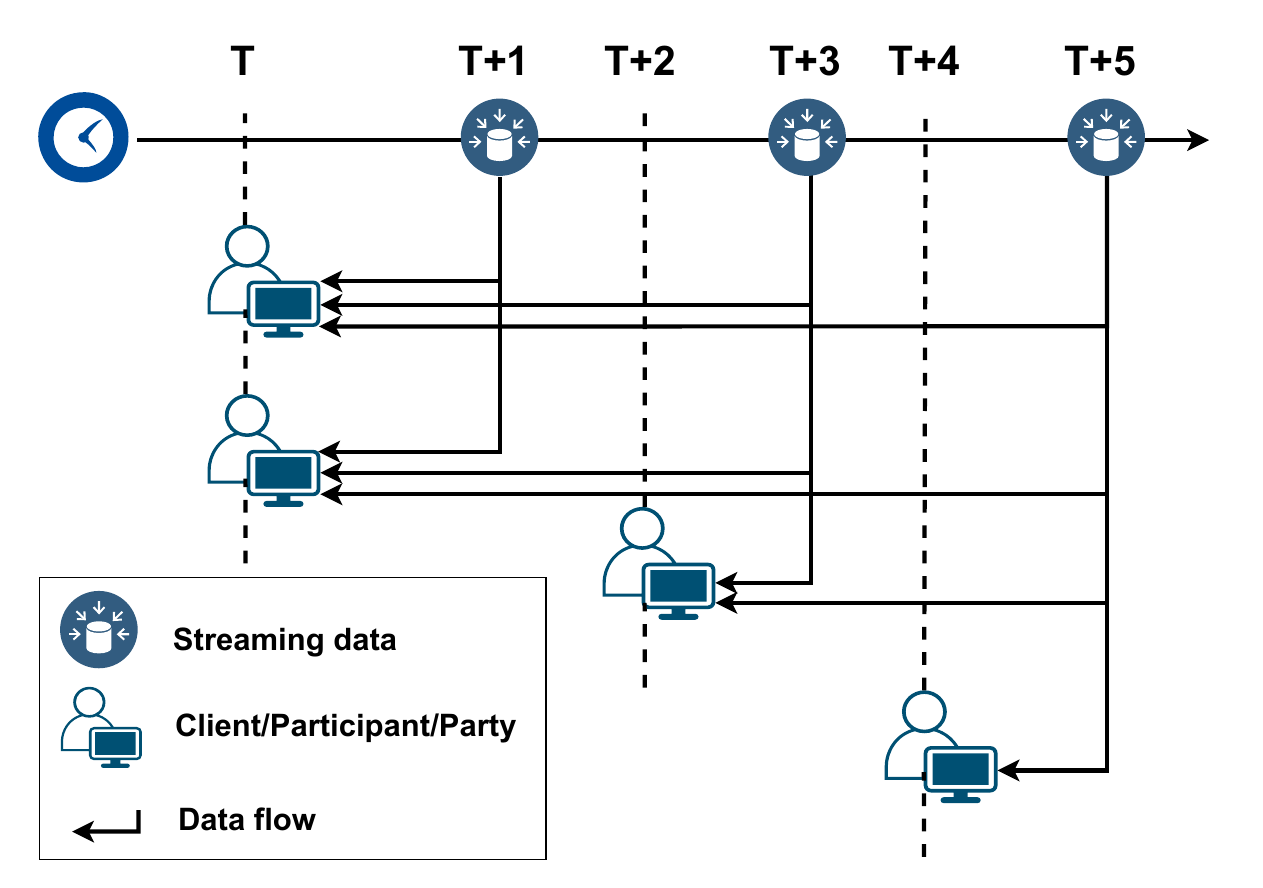}
\caption{Temporal-based splitting}
\label{fig:split-temporal}
\end{figure}

When working with temporal data (e.g. time-evolving data), we can partition the data based on time intervals, where each client gets data from different time periods. This approach reflects real-world scenarios, such as when new clients come and can only access data from a specific time onward (see Fig. \ref{fig:split-temporal}). Time-based partitioning is useful in cases such as stock price forecasting or financial risk modeling by quarters, as well as smart city management for optimizing traffic signals and predicting congestion. For instance, traffic data is collected from various sensors or cameras over time at different traffic management centers. These centers might gather data at specific times of day (morning, afternoon, evening) or during specific seasons (spring, summer, autumn, winter). Based on temporal partitioning in FL, each center can train models based on the time periods when it has the most data, while still contributing to the global model's relevance and performance.

\subsubsection{Common splitting practices in prior works} 

These data splitting strategies are often constraints imposed by an industrial context. Therefore, making synthetic real-world data distributions implies first understanding this context and then splitting the data accordingly. This captures seasonality, concept drift, and availability patterns materially affecting training. Drift and intermittent participation typically slow and destabilize convergence, especially for long local epochs or stale updates, and can degrade generalization to future windows if temporal leakage is not controlled.

From a technical perspective, the literature distributes data using several methods that span Dirichlet-based splits with varying $\alpha$ parameters (from 0.01 to 10,000), Pathological splits, feature or label shifts, nearly IID splits, and IID splits. When using a Dirichlet-based split, very small $\alpha$ (e.g., $\alpha=0.01$ in~\cite{wang2024dfrd}) leads to highly non-IID data. Conversely, very large $\alpha$ values (e.g., $\alpha=100$ in~\cite{chen2023guardhfl}) yield nearly IID partitions that often serve as a baseline to gauge improvements over highly non-IID scenarios. Moreover, pathological splits where each client sees only two or a fixed number of classes (e.g.,~\cite{xiao2023communication, ye2023personalized}) result in heavy label imbalance. Meanwhile, studies that explore feature or concept drift~\cite{bao2023optimizing, panchal2023flash} aim to mimic real-world non-stationary data distributions, adding another dimension of complexity to the FL setting.

Beyond these, the literature includes numerous application-specific variants that tailor splits to domain constraints and often fall under natural or cluster-based partitioning. In contrast, we use synthetic to denote algorithmic re-partitioning of a centralized dataset that ignores real client provenance and instead imposes controlled label or sample proportions. Within this synthetic family, the most common generator is the Dirichlet split. Other widely used synthetic label-distribution skew baselines replace $\alpha$ with simple constraints: pathological $n$-class splits (each client sees exactly $n$ labels), random-class subsets (each client draws a random label subset), and $\mu$-fraction splits (a fixed fraction $\mu$ from one dominant class with the remainder from others). These variants accentuate label-support mismatch without relying on Dirichlet draws. Table~\ref{tab:table-data-split} operationalizes these families with representative settings from the literature and emphasizes their resulting IID-ness. For Dirichlet-based rows, we report typical $\alpha$ choices and their qualitative effect on heterogeneity. For the other synthetic methods, we summarize the controlling parameter ($n$ classes, subset size, or $\mu$) and their expected effects.

{\scriptsize
\setlength{\tabcolsep}{2pt}
\renewcommand{\arraystretch}{1.06}
\begin{longtable}{|
  >{\raggedright\arraybackslash}p{0.16\linewidth}|
  >{\raggedright\arraybackslash}p{0.17\linewidth}|
  >{\raggedright\arraybackslash}p{0.31\linewidth}|
  >{\raggedright\arraybackslash}p{0.31\linewidth}|
}
\caption{Overview of data splitting strategies in the literature.}
\label{tab:table-data-split}\\
\hline
\multicolumn{1}{|c|}{\textbf{Splitting Type}} &
\multicolumn{1}{c|}{\textbf{Parameter Category}} &
\multicolumn{1}{c|}{\textbf{References}} &
\multicolumn{1}{c|}{\textbf{Notes/Applications}} \\ \hline
\endfirsthead

\hline
\multicolumn{1}{|c|}{\textbf{Splitting Type}} &
\multicolumn{1}{c|}{\textbf{Parameter Category}} &
\multicolumn{1}{c|}{\textbf{References}} &
\multicolumn{1}{c|}{\textbf{Notes/Applications}} \\ \hline
\endhead

Non-IID (Dirichlet) &
\textbf{Very small $\alpha$ ($\le 0.1$)} &
DFRD \cite{wang2024dfrd}, PFedCS \cite{wu2025pfedcs}, FedUV \cite{son2024feduv}: $\alpha=0.01$. On the Convergence of FedAvg \cite{cho2023convergence}, PFedCS \cite{wu2025pfedcs}, FedSAM/FedASAM \cite{caldarola2022fedsam}: $\alpha=0.05$. FedSAM/FedASAM \cite{caldarola2022fedsam}: $\alpha=0$. &
Extreme non-IID. DFRD focuses on data-free robustness distillation under highly skewed data. FedAvg analysis at $\alpha=0.05$ with cyclic client participation (practical FL). 
\\ \hline

Non-IID (Dirichlet) &
\textbf{Small $\alpha$ ($\approx 0.1$--$0.3$)} &
Flow \cite{panchal2024flow}, Flash \cite{panchal2023flash}, DFRD \cite{wang2024dfrd}, DELTA \cite{wang2024delta}, IBA \cite{nguyen2024iba}, Lockdown \cite{huang2024lockdown}, FedMRUR \cite{an2023federated}, FedGELA \cite{fan2024federated}, FedCS \cite{hao2025fedcs}, PFedCS \cite{wu2025pfedcs}, FedConcat \cite{diao2024fedconcat}, SphereFed \cite{dong2022spherefed}, FedRIR \cite{huang2025fedrir}, FedLDR \cite{zhu2025fedldr}: $\alpha=0.1$. Lockdown, Fed-CBS \cite{zhang2023fed}, FedMRUR, Fed-CO2 \cite{cai2024fed}: $\alpha=0.2$. FedCR \cite{zhang2023fedcr}, Fed-CO2, FedDC \cite{gao2022feddc}, FedDyn \cite{acar2021feddyn}: $\alpha=0.3$. &
Strongly non-IID. Flash/Flow introduce concept drift or per-instance personalization. Lockdown: partial poisoning at $\alpha\approx0.2$. Fed-CBS: class imbalance. FedGELA: small and large $\alpha$ to model class-disjoint data. FedCS studies coreset selection under $\alpha=0.1$.
\\ \hline

Non-IID (Dirichlet) &
\textbf{Medium $\alpha$ ($\approx 0.3$--$1$)} &
DELTA, Lockdown, IBA, Fed-CBS, FedGELA, FedRN \cite{kim2022fedrn}, FedELC \cite{jiang2024fedelc}, FedConcat \cite{diao2024fedconcat}, FedBABU \cite{oh2022fedbabu}, SphereFed \cite{dong2022spherefed}, FedLTN \cite{mugunthan2022fedltn}, FedLDR \cite{zhu2025fedldr}: $\alpha=0.5$. 
Flash, Flow, FedMRUR, FedDC, FedDyn, FedLDR: $\alpha=0.6$. FedLTN \cite{mugunthan2022fedltn}: $\alpha=0.7$. Lockdown: $\alpha=0.8$. Dynamic Personalized FL \cite{yang2023dynamic}, FedMR \cite{hu2024aggregation}, CASA \cite{liu2024boyi}, FedSAC \cite{wang2024fedsac}, GuardHFL \cite{chen2023guardhfl}, DFRD, FedUV, FedELC, FedBABU: $\alpha=1$. &
Moderately non-IID. IBA uses $\alpha=0.5$ for backdoor tasks. FedRN/FedELC combine Dirichlet heterogeneity with noisy-label settings. $\alpha=1$ is often still non-IID but less skewed; widely used as a ``moderate'' baseline.
\\ \hline

Non-IID (Dirichlet) &
\textbf{Large $\alpha$ ($\approx 1$--$100$)} &
FedSAC \cite{wang2024fedsac}: $\alpha=2,3$. Dynamic Personalized FL \cite{yang2023dynamic}, FedLDR \cite{zhu2025fedldr}: $\alpha=10$. &
As $\alpha$ grows beyond 1, distributions approach IID.
\\ \hline

Nearly IID (Dirichlet) &
\textbf{Very large $\alpha$ ($\ge 100$)} &
Dynamic Personalized FL \cite{yang2023dynamic}, GuardHFL \cite{chen2023guardhfl}, FedSAM/FedASAM \cite{caldarola2022fedsam}, FedLDR \cite{zhu2025fedldr}: $\alpha=100$. FedSAM/FedASAM \cite{caldarola2022fedsam}: $\alpha=1000$. FedGELA \cite{fan2024federated}: $\alpha=10000$. &
Effectively near-IID at $\alpha\ge100$; extreme near-IID at $10{,}000$. FedGELA merges partial class-disjoint data with huge $\alpha$.
\\ \hline

Non-IID (Pathological) &
\textbf{$n=1, 2$ classes or shards} &
Maverick \cite{wang2025maverick} (1 class/client). Communication-Efficient Fed Hypergradient \cite{xiao2023communication}, FedUFO \cite{zhang2021fedufo}, FedRN \cite{kim2022fedrn} (2 shards each). Personalized FL \cite{ye2023personalized} (some runs). Anchor Sampling \cite{wu2023anchor}, Towards personalized FL \cite{wang2024towards}, PFedCS \cite{wu2025pfedcs}, FedConcat \cite{diao2024fedconcat}, FedLTN \cite{mugunthan2022fedltn}, FedACA \cite{dong2026fedaca}, FedDSE \cite{wang2024feddse} (2 classes/client) &
LEAF-style split: each client sees exactly 1 or sometimes 2 classes or shards (shards are usually very small with limited samples); very strong label imbalance; synthetic cluster-based split.
\\ \hline

Non-IID (Pathological) &
\textbf{$n=3,4,5$ classes or shards} &
PFedCS \cite{wu2025pfedcs}, FedConcat \cite{diao2024fedconcat}, FedDSE \cite{wang2024feddse}: 3 or 4 classes each. FedPHP \cite{li2021fedphp}, FedDSE \cite{wang2024feddse}: 5 classes each. FedRN \cite{kim2022fedrn}: 5 shards each. &
Intermediate pathological label-skew. FedPHP is explicitly clear using 100 clients, 5 classes each with 400 train and 100 test samples/client on CIFAR10.
\\ \hline

Non-IID (Pathological) &
\textbf{$n=10$ classes or shards} &
Fed-CO2 \cite{cai2024fed}, FedMRUR \cite{an2023federated}, PFedCS \cite{wu2025pfedcs}, FedACA \cite{dong2026fedaca}, FedDSE \cite{wang2024feddse}: 10 classes/client. &
Each client has exactly 10 classes, still partial coverage of the label set. Synthetic cluster-based split.
\\ \hline

Non-IID (Pathological) &
\textbf{$n=20,30, 40,80$ classes or shards} &
Personalized FL \cite{ye2023personalized}, FedPHP \cite{li2021fedphp}: 20 classes each. PFedCS \cite{wu2025pfedcs}: $n=20, 30,40$. FedMRUR \cite{an2023federated}: $n=40,80$. &
20--80 classes or shards per client. Higher $n$ means less skew, but still forced partial class coverage.
\\ \hline
Non-IID (Feature Shift) &
\textbf{Rotations, Concept Drift} &
FedCOLLAB \cite{bao2023optimizing}: rotations ($\pm25^\circ,\pm155^\circ$). Flash \cite{panchal2023flash}: concept drift (0.1--1.0). &
Feature shift instead of label imbalance (e.g., rotated images). Flash also updates the distribution over time. Synthetic split to model natural phenomena.
\\ \hline

Non-IID (Label Shift) &
\textbf{Partial overlap of labels} &
FedCOLLAB \cite{bao2023optimizing}: partial label overlap (client 0 vs. others). &
Label proportions differ substantially across clients; some are fully aligned or fully disjoint. Synthetic split to model natural phenomena.
\\ \hline

Non-IID (Random Class) &
\textbf{Random subset of classes} &
FedCR \cite{zhang2023fedcr}: random 5 or 15 classes/client. &
Class subset chosen randomly per client, without Dirichlet. Application-specific synthetic split.
\\ \hline

Non-IID ($\mu$-Fraction) &
\textbf{$\mu=0.3$ main class + leftover} &
No One Idles \cite{zhang2023no}. &
A fraction $\mu$ from a single “primary” class; the rest from others. Application-specific synthetic split.
\\ \hline

Non-IID/IID (Mixed/Hybrid) &
\textbf{Different parameter kinds} &
DoCoFL \cite{dorfman2023docofl}: 10\% overall IID + 90\% IID from 2 classes. FedIns \cite{feng2023fedins}: intra-client and inter-client heterogeneity.&
Combine methods to model a target scenario (real-world or partial IID + non-IID). Application-specific synthetic split.
\\ \hline

IID (Divide uniformly) &
\textbf{No parameter} &
No One Idles \cite{zhang2023no}, Lockdown \cite{huang2024lockdown}, FedDC \cite{gao2022feddc}&
Uniformly split the dataset over clients.
\\ \hline
\end{longtable}
}

These results justify Table~\ref{tab:table-data-split-proof} by linking observed accuracy and convergence behavior to the underlying split mechanism, client participation, and model architecture. The table summarizes reported outcomes, making two primary drivers explicit: the distribution of data across clients (split type and its parameters such as Dirichlet $\alpha$ or the number of classes/shards per client) and the effective client participation ratio (denoted as joined/total clients in each round).

For Dirichlet splits, the dominant factor is the concentration parameter $\alpha$. Very small values such as $\alpha\leq 0.1$ create extreme label-distribution skew, where each client observes a highly concentrated subset of labels. This setting often leads to low or unstable accuracy, especially on harder datasets such as CIFAR-100 and Tiny-ImageNet, although strong personalization or specialized methods can still obtain high values on easier datasets or under full participation. Small $\alpha$ values around $0.1$--$0.3$ remain strongly non-IID, but the reported results show a wider range because they include different methods, models, and participation ratios. Medium $\alpha$ values around $0.3$--$1$ reduce the severity of label skew and usually improve stability within the same dataset/model family. Large and very large $\alpha$ values move the partition closer to IID, and the corresponding rows show stronger performance, especially on Fashion-MNIST, CIFAR-10, and near-IID Dirichlet baselines. However, it is noted that these rows should not be compared blindly across datasets: for example, CIFAR-100 and Tiny-ImageNet remain harder even when the split is less skewed.

Pathological splits expose a higher-level, tougher form of heterogeneity, where clients receive a fixed number of classes or shards. The table shows that the most restrictive settings, such as $n=1$ or $n=2$ classes (or shards) per client, produce severe label imbalance and can sharply reduce performance, especially on datasets with many classes, such as CIFAR-100 and Tiny-ImageNet. Increasing the number of local classes or shards, for example, from $n=2$ to $n=10$ or $n=20$--$80$, generally weakens the heterogeneity as each client covers a larger portion of the label space, facilitating reaching higher accuracy. The improvement is clearest when comparing results within the same dataset and model family.

Feature-shift and label-shift rows capture heterogeneity without any described numbers like Dirichlet $\alpha$ or the number of classes per client. In feature-shift settings, such as rotations or concept drift, clients may have the same label space but different input distributions or time-varying concepts. In label-shift settings with partial label overlap, the client label supports are manually overlapped or separated rather than sampled by a standard Dirichlet or fixed-class rule. These settings can degrade aggregation even when the class count alone appears reasonable, because local objectives differ in either input space, label support, or temporal distribution.

Model architecture is another confounding factor. Differences among generic CNN entries can partly reflect architectural design and hyperparameter tuning choices rather than just the underlying split. For this reason, rows that report a standard and fixed model, such as ResNet18, are more informative for isolating the effect of data partitioning and participation. Notably, the table reveals that more advanced architectures, such as Vision Transformers (\textit{ViT-B/16}), exhibit strong resilience to non-IID data, even on data like CIFAR-100 or Tiny-ImageNet. For example, it can achieve 84.8\% in mixed settings and 75.25--77.06\% on CIFAR-100 even under a highly skewed Dirichlet. Nevertheless, within any fixed model family and dataset, the dominant trend remains absolute: harsher non-IID splits and lower participation ratios reduce accuracy and can prolong convergence. 

Client participation provides important context for Table~\ref{tab:table-data-split-proof}, but it is not a standalone explanation for accuracy differences. Lower participation, can increase update variance and slow convergence because each round observes only a partial view of the global client population. In contrast, higher participation, usually gives a more stable estimate of the global update direction. This holds true only under the same controlled experimental design.

\begin{figure}[h!]
\centering
\includegraphics[width=0.75\textwidth]{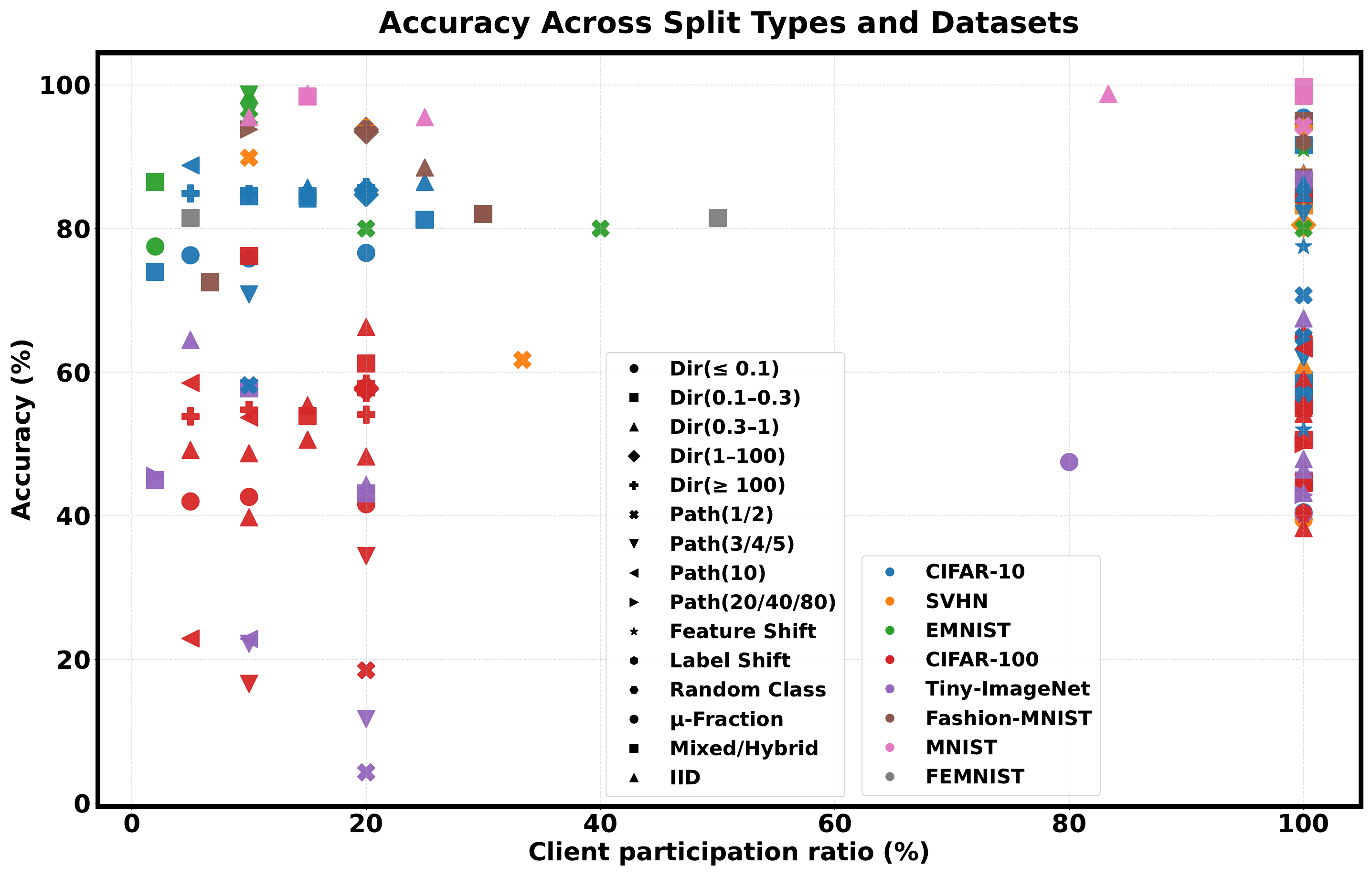}
\caption{2D scattered accuracy by split and datasets.}
\label{fig:split-2d-part-ratio}
\end{figure}

\begin{figure}[h!]
\centering
\includegraphics[width=0.78\textwidth]{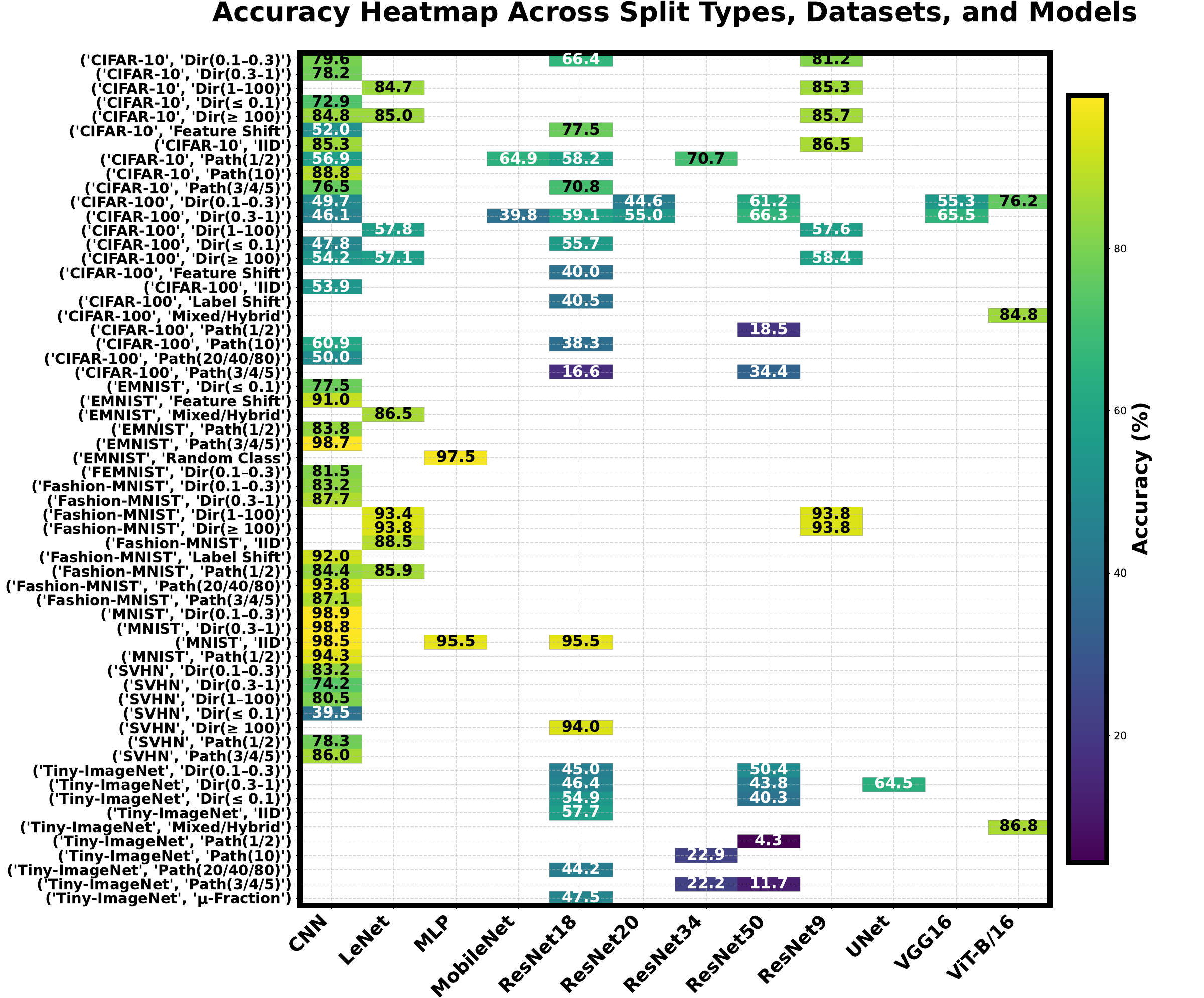}
\caption{Accuracy heatmap Across split, datasets, and models.}
\label{fig:split-2d-all-avg}
\end{figure}

\begin{figure}[h!]
\centering
\includegraphics[width=0.75\textwidth]{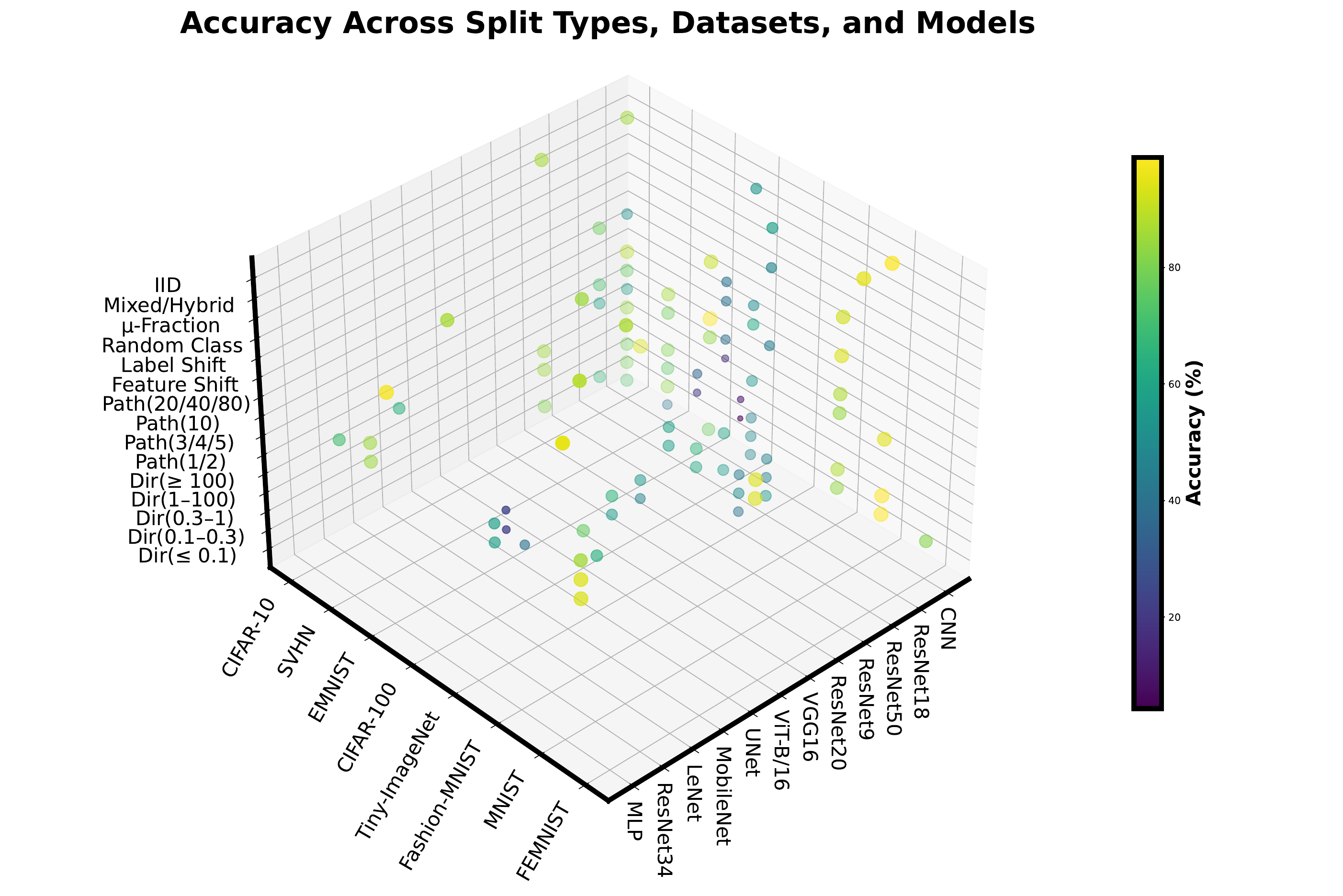}
\caption{3D scattered accuracy across split, datasets, and models.}
\label{fig:split-3d-all-avg}
\end{figure}

Figure~\ref{fig:split-2d-part-ratio} complements Table~\ref{tab:table-data-split-proof} by visualizing how the client participation ratio interacts with split severity. Figure~\ref{fig:split-2d-all-avg} further aggregates the table-level evidence into an average accuracy heatmap across splitting types, datasets, and models. See Figure~\ref{fig:split-3d-all-avg} for a 3D view of this. Together, the table and figures show both the paper-level variability and the consistent average trend. In which not one factor decides all, but how heterogeneous the data and partial participation jointly shape the final accuracy and convergence behavior of FL methods.

{\scriptsize
\setlength{\tabcolsep}{2pt}
\renewcommand{\arraystretch}{1.08}
\setlength{\LTcapwidth}{\textwidth}

\begin{longtable}{|
  >{\raggedright\arraybackslash}p{0.15\linewidth}|
  >{\raggedright\arraybackslash}p{0.19\linewidth}|
  >{\raggedright\arraybackslash}p{0.59\linewidth}|
}
\caption{Overview of data splitting strategies (continue from Table \ref{tab:table-data-split}). Convergence/performance proof is displayed in the format of \textbf{Dataset:} Accuracy\%-[joined/total clients]-Model.}
\label{tab:table-data-split-proof}\\
\hline
\multicolumn{1}{|c|}{\textbf{Splitting Type}} &
\multicolumn{1}{c|}{\textbf{Parameter Category}} &
\multicolumn{1}{c|}{\textbf{Convergence/Performance Proof}} \\ \hline
\endfirsthead

\hline
\multicolumn{1}{|c|}{\textbf{Splitting Type}} &
\multicolumn{1}{c|}{\textbf{Parameter Category}} &
\multicolumn{1}{c|}{\textbf{Convergence/Performance Proof}} \\ \hline
\endhead

\hline

Non-IID (Dirichlet) &
\begin{tabular}[c]{@{}l@{}}
\textbf{Very small $\alpha$}
\textbf{($\le 0.1$)}
\end{tabular}
&
\begin{tabular}[c]{@{}l@{}}
- \textbf{CIFAR-10:} 40--41\%-[10/10]-CNN; 93.58--97.30\%-[20/20]-CNN;\\
\quad 76.12--76.44\%/75.51--76.16\%/76.36--76.86\%-[5, 10, 20/100]-CNN;\\
- \textbf{SVHN:} 39--40\%-[10/10]-CNN\\
- \textbf{EMNIST:} 75--80\%-[10/500]-CNN\\
- \textbf{CIFAR-100:} 55.7\%-[10/10]-ResNet18; 59.39--70.20\%-[20/20]-CNN;\\
\quad 42.01\%/42.64\%/41.62\%-[5, 10, 20/100]-CNN\\
- \textbf{Tiny-ImageNet:} 40.3\%-[10/10]-ResNet50; 49.55--60.26\%-[20/20]-ResNet18
\end{tabular}
\\ \hline

Non-IID (Dirichlet) &
\begin{tabular}[c]{@{}l@{}}
\textbf{Small $\alpha$}
\textbf{($\approx 0.1$--$0.3$)}
\end{tabular}
&
\begin{tabular}[c]{@{}l@{}}
- \textbf{CIFAR-10:} 57.7\%/84--85\%/91.56--91.64\%-[40/40]/[10, 15/100]/[20/20]-CNN;\\
\quad 79.5--83\%-[10/40]-ResNet9; 58.88\%/$\sim$71--77\%-[10/10]/[10/500]-ResNet18\\
- \textbf{Fashion-MNIST:} $\sim$82\%-[36/120, 60/200]-CNN; 72--73\%-[20/300]-CNN;\\
\quad 92.53--97.51\%-[20/20]-CNN; 84.5\%-[40/40]-CNN\\
- \textbf{MNIST:} 98.40\%/98.51\%/99.69\%-[15/100]/[100/100]/[20/20]-CNN\\
- \textbf{SVHN:} 83.2\%-[40/40]-CNN\\
- \textbf{FEMNIST:} 81--82\%-[10/20--200]-CNN\\
- \textbf{Tiny-ImageNet:} $\sim$45\%/43.22--46.60\%-[10/500]/[20/20]-ResNet18;\\
\quad 43.1\%/56.83--58.67\%-[40/200]/[10/100]-ResNet50\\
- \textbf{CIFAR-100:} 34.5--54.86\%/50.60\%/53.91\%-[100/100]/[20/20]/[15/100]-CNN; \\
\quad 44.6\%-[100/100]-ResNet20; 55.33\%-[100/100]-VGG16; \\
\quad 61.2\%-[40/200]-ResNet50; 75.25--77.06\%-[10/100]-ViT-B/16
\end{tabular}
\\ \hline

Non-IID (Dirichlet) &
\begin{tabular}[c]{@{}l@{}}
\textbf{Medium $\alpha$}
\textbf{($\approx 0.3$--$1$)}
\end{tabular}
&
\begin{tabular}[c]{@{}l@{}}
- \textbf{Tiny-ImageNet:} 43.2\%/44.3\%-[10/10]/[40/200]-ResNet50;\\
\quad 46.44\%-[20/20]-ResNet18; 64--65\%-[10/200]-UNet\\
- \textbf{MNIST:} $\sim$98.5--99\%-[100/100, 15/100, 100/120]-CNN;\\
- \textbf{Fashion-MNIST:} 87.7\%-[40/40]-CNN\\
- \textbf{SVHN:} 58--64\%-[50/50]-CNN; 87.5\%-[40/40]-CNN\\
- \textbf{CIFAR-10:} 64.2\%-[40/40]-CNN; 84.77--85.64\%-[15, 100/100]-CNN\\
- \textbf{CIFAR-100:} 37.4--39.2\%/-[100/100]-CNN; 34.9--44.7\%-[10/100]-MobileNet;\\
\quad 48.27\%/48.72\%/49.17\%-[20, 10, 5/100]-CNN;\\
\quad 54.3--55.8\%-[100/100]-ResNet20; 59.1\%-[10/10]-ResNet18;\\
\quad 66.3\%-[40/200]-ResNet50; 65--66\%-[100/100]-VGG16
\end{tabular}
\\ \hline

Non-IID (Dirichlet) &
\textbf{Large $\alpha$ ($\approx 1$--$100$)}
&
\begin{tabular}[c]{@{}l@{}}
- \textbf{SVHN:} 79--82\%-[10/10]-CNN\\
- \textbf{Fashion-MNIST:} $\sim$93.8\%/93.24--93.65\%-[4/20]-ResNet9/LeNet\\
- \textbf{CIFAR-10:} 85--85.68\%/84.4--85.01\%/[4/20]-ResNet9/LeNet\\
- \textbf{CIFAR-100:} 57.03--58.19\%/57.72--57.93\%/[4/20]-ResNet9/LeNet
\end{tabular}
\\ \hline

Nearly IID (Dirichlet) &
\begin{tabular}[c]{@{}l@{}}
\textbf{Very large $\alpha$}
\textbf{($\ge 100$)}
\end{tabular}
&
\begin{tabular}[c]{@{}l@{}}
- \textbf{SVHN:} $\sim$94\%-[10/10, 10/50]-ResNet18\\
- \textbf{Fashion-MNIST:} $\sim$93.8\%-[4/20]-ResNet9/LeNet\\
- \textbf{CIFAR-10:} 84.88\%/84.80\%/84.79\%-[5, 10, 20/100]-CNN;\\
\quad 85.73\%/85.03\%-[4/20]-ResNet9/LeNet\\
- \textbf{CIFAR-100:} 53.86\%/54.79\%/54.10\%-[5, 10, 20/100]-CNN;\\
\quad 58.4\%/57.07\%-[4/20]-ResNet9/LeNet
\end{tabular}
\\ \hline

Non-IID (Pathological) &
\textbf{$n=1,2$ classes or shards}
&
\begin{tabular}[c]{@{}l@{}}
- \textbf{EMNIST:} $\sim$80\%-[20, 40, 100/100]-CNN; 95.34\%-[10/100]-CNN\\
- \textbf{SVHN:} 61.72\%/83.4\%/89.87\%-[4/12]/[40/40]/[10/100]-CNN\\
- \textbf{MNIST:} 92.94--95.61\%-[100/100]-CNN\\
- \textbf{Fashion-MNIST:} 84.4\%-[40/40]-CNN; 85.9\%-[50/50]-LeNet\\
- \textbf{CIFAR-10:} 56.9\%-[40/40]-CNN; 58.19\%-[10/100]-ResNet18;\\
\quad 64.9\%/70.7\%-[50/50]-MobileNet/ResNet34\\
- \textbf{CIFAR-100:} 18.5\%-[40/200]-ResNet50\\
- \textbf{Tiny-ImageNet:} 4.3\%-[40/200]-ResNet50
\end{tabular}
\\ \hline

Non-IID (Pathological) &
\textbf{$n=3,4,5$ classes or shards}
&
\begin{tabular}[c]{@{}l@{}}
- \textbf{CIFAR-10:} 62.0\%-[40/40]-CNN; 85.49/82.06\%-[20/20]-CNN;\\
\quad 70.82\%-[10/100]-ResNet18\\
- \textbf{SVHN:} 86.0\%-[40/40]-CNN\\
- \textbf{Fashion-MNIST:} 87.1\%-[40/40]-CNN\\
- \textbf{EMNIST:} 98.65\%-[10/100]-CNN\\
- \textbf{CIFAR-100:} 16.61\%-[10/100]-ResNet18; 34.4\%-[40/200]-ResNet50\\
- \textbf{Tiny-ImageNet:} 11.7\%-[40/200]-ResNet50; 22.19\%-[10/100]-ResNet34
\end{tabular}
\\ \hline

Non-IID (Pathological) &
\textbf{$n=10$ classes or shards}
&
\begin{tabular}[c]{@{}l@{}}
- \textbf{CIFAR-10:} 88.79\%-[5/100]-CNN\\
- \textbf{CIFAR-100:} 58.50\%/63.29\%-[5/100]/[20/20]-CNN; \\
\quad 22.93\%/53.69\%-[10/200]/[10/100]-ResNet18\\
- \textbf{Tiny-ImageNet:} 22.88\%-[10/100]-ResNet34
\end{tabular}
\\ \hline

Non-IID (Pathological) &
\textbf{$n=20,30,40,80$ classes or shards}
&
\begin{tabular}[c]{@{}l@{}}
- \textbf{Fashion-MNIST:} 93.8\%-[10/100]-CNN\\
- \textbf{Tiny-ImageNet:} 45.42--45.71\%-[10/500]-ResNet18;\\
\quad 38.92--46.93\%-[20/20]-ResNet18\\
- \textbf{CIFAR-100:} $\sim$50\%-[20/20]-CNN
\end{tabular}
\\ \hline

Non-IID (Feature Shift) &
\textbf{Rotations, Concept Drift parameters}
&
\begin{tabular}[c]{@{}l@{}}
- \textbf{EMNIST:} 90--92\%-[10/10]-CNN\\
- \textbf{CIFAR-100:} $\sim$40\%-[10/10]-ResNet18\\
- \textbf{CIFAR-10:} 76--79\%-[10/10]-ResNet18; $\sim$52\%-[20/20]-CNN
\end{tabular}
\\ \hline

Non-IID (Label Shift) &
\shortstack[l]{\textbf{Partial overlap of}\\ \textbf{labels}}
&
\begin{tabular}[c]{@{}l@{}}
- \textbf{CIFAR-100:} 40--41\%-[20/20]-ResNet18\\
- \textbf{Fashion-MNIST:} $\sim$92\%-[20/20]-CNN
\end{tabular}
\\ \hline

Non-IID (Random Class) &
\textbf{Random subset of classes}
&
\begin{tabular}[c]{@{}l@{}}
- \textbf{EMNIST:} 97.47\%-[10/100]-MLP
\end{tabular}
\\ \hline

Non-IID ($\mu$-Fraction) &
\textbf{$\mu=0.3$ main class + leftover}
&
\begin{tabular}[c]{@{}l@{}}
- \textbf{Tiny-ImageNet:} 47--48\%-[16/20(4-dropout)]-ResNet18
\end{tabular}
\\ \hline

Non-IID/IID (Mixed/Hybrid) &
\textbf{Different kinds of parameters}
&
\begin{tabular}[c]{@{}l@{}}
- \textbf{EMNIST:} 86--87\%-[20/1000]-LeNet\\
- \textbf{CIFAR-100:} 84.83\%-[5/5]-ViT-B/16\\
- \textbf{Tiny-ImageNet:} 86.79\%-[5/5]-ViT-B/16
\end{tabular}
\\ \hline

IID (Divide uniformly) &
\textbf{No parameter}
&
\begin{tabular}[c]{@{}l@{}}
- \textbf{MNIST:} 93--98\%-[10/40, 10/100]-ResNet18/MLP;\\
\quad 98.45/98.47\%-[100, 15/100]-CNN\\
- \textbf{CIFAR-10:} 85--88\%-[10/40]-ResNet9; 86.18/85.71\%-[100, 15/100]-CNN;\\
\quad 84.93/84.19\%-[500, 75/500]-CNN\\
- \textbf{CIFAR-100:} 55.52/55.40\%-[100, 15/100]-CNN;\\
\quad 54.25/50.61\%-[500, 75/500]-CNN\\
- \textbf{Tiny-ImageNet:} 47.91\%/67--68\%-[20/20]-ResNet18\\
- \textbf{Fashion-MNIST:} 88--89\%-[10/40]-LeNet
\end{tabular}
\\ \hline
\end{longtable}
}

\subsection{Assisting convergence in non-IID distributions}
Performance and convergence typically deteriorate as data becomes more non-IID across clients~\cite{hsu2019measuring, bettinelli2024discovering}, underscoring the importance of carefully choosing and tuning these splitting strategies. To meet this challenge, several approaches have emerged. The first solution focuses on the data itself. When data is identified as non-IID, augmentation techniques can be applied to rebalance the distributions. To illustrate, an auto-encoder can be used to generate new data of missing samples \cite{wen2022communication, li2024feature}. Nevertheless, in most real-world settings, clients do not reveal their local data distributions, which complicates even the detection of heterogeneity. This leads to a second solution, which mitigates the impact of non-IID data on model optimizations without generating new samples. The category encompasses several techniques, including: 


\textit{Local fine-tuning:} helps mitigate the impact of \textit{non-IIDness} on client models by adapting the model received from the server with clients' local data. While this method helps preserve clients' model performance, it does not improve the optimization of the global model, which remains affected by clients' non-IID data.

\textit{Personalization layer:} splitting the model into two parts: personalized layers and base layers. Like standard FL, base layers are sent to a server for aggregation. But each client trains personalized layers on their own data \cite{arivazhagan2019federated, liang2020think}, ensuring that clients maintain an adapted model to their data distribution. This technique also optimizes a global model more robust to clients' \textit{non-IIDness}.  

\textit{Multi-task learning:} gathers clients with similar data distributions to learn together. Each client group has its own cluster model that is shared among cluster members, enabling clients to better fit to their data. This concept was first introduced by \textit{Sattler et al.} as \textit{Clustered FL} \cite{sattler2020clustered, islam2024fedclust}. However, this technique only mitigates the impact of \textit{non-IIDness} on the client and cluster models.

\textit{Knowledge distillation:} transfers information at the prediction level. Each client trains its model locally and then shares only the outputs on a small public or generated dataset. The server aggregates these labels to train a global model that captures the consensus knowledge of all clients. By aligning models through outputs rather than parameters, knowledge distillation reduces the bias caused by heterogeneous data. It enables the global model to generalize better across diverse clients while preserving data privacy \cite{zhao2024data, sun2025fedagent}. However, this technique requires having data on the server side, which is not always compatible with real-world scenarios.

\section{Data vulnerabilities}
\label{sec:security}

This section discusses data-related vulnerabilities in FL, spanning from common poisoning and inference attacks to less discussed evasion and often overlooked leakage from replay buffers. We first summarize each attack via \textit{``The five Ws and How''} in Table~\ref{tab:sec-summary}, then synthesize defending methods and their trade-off evaluations in Table~\ref{tab:vul-tradeoff}.

It is worth noting that this section is not intended as a direct benchmark comparison with the non-adversarial convergence studies discussed earlier. The vulnerability-related works considered here are typically evaluated under specific attack models, poisoning ratios, adversarial assumptions, and defense settings. Therefore, we interpret their reported convergence behavior as evidence under adversarial data conditions, while focusing on how defense mechanisms affect clean and under-attack utility.

We assess defenses using a Pareto~\cite{miettinen1999nonlinear, deb2011multi} lens to make trade-offs explicit. In this view, a method Pareto-dominates another if it is at least as good as, or equal to, on every relevant axis and strictly better than on at least one. We focus on two axes of utility: (i) \textit{clean utility} (standard accuracy or convergence when no attack is present) and (ii) \textit{under-attack utility} (robust accuracy, attack-success reduction, or any related metrics). The Pareto frontier is the set of non-dominated methods -- those for which no alternative yields both equal-or-higher clean utility and equal-or-higher under-attack utility simultaneously. Positioning defenses on this frontier helps to see whether gains in robustness come at measurable costs to clean performance. 

Finally, to complement the method-level trade-off summary in Table~\ref{tab:vul-tradeoff}, we provide Table~\ref{tab:defense-guidelines}, which distills practical defense-selection guidelines by attack scenario under IID and non-IID regimes. Since the degree of heterogeneity is not explicitly reported in all surveyed works, we use ``non-IID'' in a broad range of skewness, from large to very small Dirichlet $\alpha$ and shard-based splits with few classes per client, as summarized in Table~\ref{tab:table-data-split}. With caution, our ``Notes'' column clarifies the principal assumptions and constraints regarding these guidelines, and highlights cases where recommendations may be sensitive to stronger heterogeneity. Ultimately, this table can serve as actionable guidance for selecting defenses across common attack scenarios and data distributions.

\begin{table}[H]
\caption{Common data-related vulnerabilities on FL: 5W and How?}
\label{tab:sec-summary}
\begin{threeparttable}
\def\arraystretch{1.2}
\resizebox{\textwidth}{!}{
\begin{tabular}{|l|c|c|l|l|l|}
\hline
\multicolumn{1}{|c|}{\textbf{What}} & \textbf{Who} & \textbf{When} & \multicolumn{1}{c|}{\textbf{Where}} & \multicolumn{1}{c|}{\textbf{Why}}                                                                                     & \multicolumn{1}{c|}{\textbf{How}}                                                                 \\ \hline
Data Poisoning                                                                         & C                                                                          & T                                                                            & Model integrity                                                                                & \begin{tabular}[c]{@{}l@{}}- Misclassify specific input yet maintain \\good performance on others (targeted)\\ - Degrade overall model performance \\(untargeted)\end{tabular} & \begin{tabular}[c]{@{}l@{}}- Flip labels from correct to different one \\ (Label flipping)\\ - Create a backdoor with hidden triggers\\ (Backdoor)\end{tabular} \\ \hline
Evasion                                                                                & C                                                                          & I                                                                            & Model decision boundary                                                                        & \begin{tabular}[c]{@{}l@{}}Cause learned model to make incorrect \\decisions \end{tabular}                                                                                                                          & Add subtle change to infererence input                                                                                                                 \\ \hline
Membership inference                                                                   & A, E                                                                       & T, I                                                                            & Data privacy                                                                                   & \begin{tabular}[c]{@{}l@{}}Find out if the data was included in \\training data \end{tabular}                                          & \begin{tabular}[c]{@{}l@{}}Infer membership based on model prediction\\ output (black-box) or the change of gradients \\(white-box)\end{tabular}        \\ \hline
\begin{tabular}[c]{@{}l@{}}Unintended leakage\\ (replay's vulnerability)\end{tabular}  & C                                                                          & T                                                                            & Data privacy                                                                                   & \begin{tabular}[c]{@{}l@{}}Extract historical data and apply malicious \\ actions later to it\end{tabular}                                                                & \begin{tabular}[c]{@{}l@{}}Exploit replay buffer to retrieve historical \\trained data \end{tabular}                             \\ \hline
\end{tabular}
}
\vspace{0.1cm}
\begin{tablenotes}[flushleft]
\scriptsize{
\item \textbf{What:} attack name | \textbf{Who:} possible attacker. Client (C), Aggregator (A), or External actor (E)
\item \textbf{When:} attack occurs at which phase of the learning - Training (T)  or Inference (I) | \textbf{Where:} the aspect being impacted
\item \textbf{Why:} short description of the attack's purpose | \textbf{How:} short description of the attack's approach or method
}
\end{tablenotes}
\end{threeparttable}
\end{table}

\subsection{Data poisoning} Data poisoning is a common attack in ML in general~\cite{zhao2018data} and FL~\cite{sun2021data, tolpegin2020data}. To poison the data, attackers intentionally inject malicious data samples into the training set, corrupting the model and potentially leading to biased decision-making. In FL settings, the possibility of this attack is even much higher when models are trained decentrally across multiple clients, each with its own local dataset. Malicious clients can insert poisoned data during the local training process, and these poisoned model updates are then aggregated into the global model. Due to privacy constraints and the non-IID nature of FL, it is more challenging to detect abnormal or malice during aggregation. 

\begin{figure}[h!]
\centering
\includegraphics[width=0.6\textwidth]{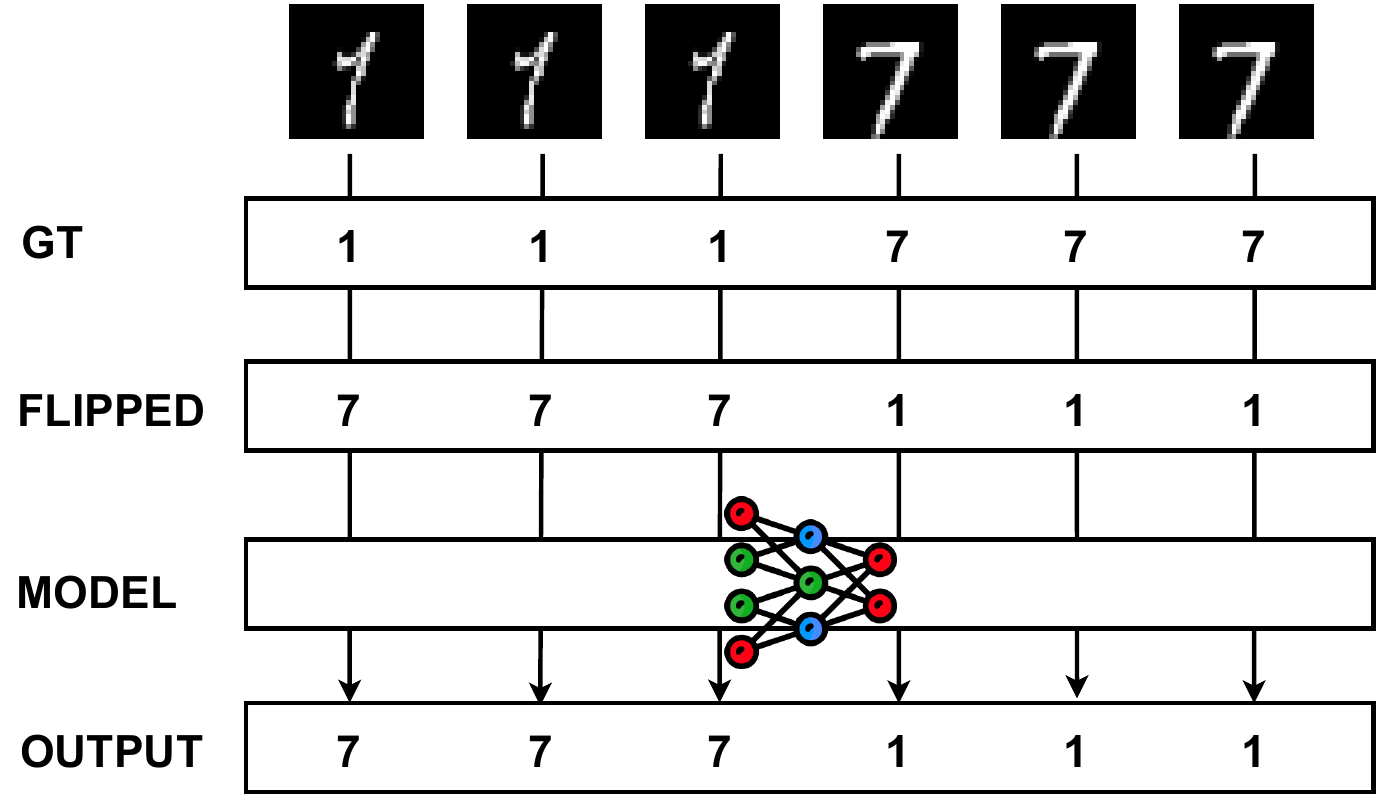}
\caption{Label-flipping attack}
\label{fig:sec-poisoning-flipping}
\end{figure}

Researchers normally categorize the poisoning attacks as targeted and untargeted attacks, in which the targeted attacks focus on causing the model to misclassify specific data points, yet maintain good performance on other inputs. On the other hand, untargeted poisoning aims to degrade the overall model performance. Each of these types of poisoning also comes along with different techniques. Among these, label flipping~\cite{cao2019understanding, jiang2023data, shen2023privacy, tolpegin2020data} and backdoor attacks~\cite{gong2022backdoor, sun2019can, wang2020attack, bagdasaryan2020backdoor} are most widely discussed.
In label-flipping attacks (see Fig. \ref{fig:sec-poisoning-flipping}), the adversary changes (or flips) all the labels of certain training samples to different ones (e.g. class 1 to 7, or class 4 to 9 in the MNIST dataset), misleading the model to learn the wrong mappings between inputs and outputs. This kind of attack is simple to execute yet effective since the attackers just need to work directly with the label without knowing the underlying model. 

\begin{figure}[h!]
\centering
\includegraphics[width=.6\textwidth]{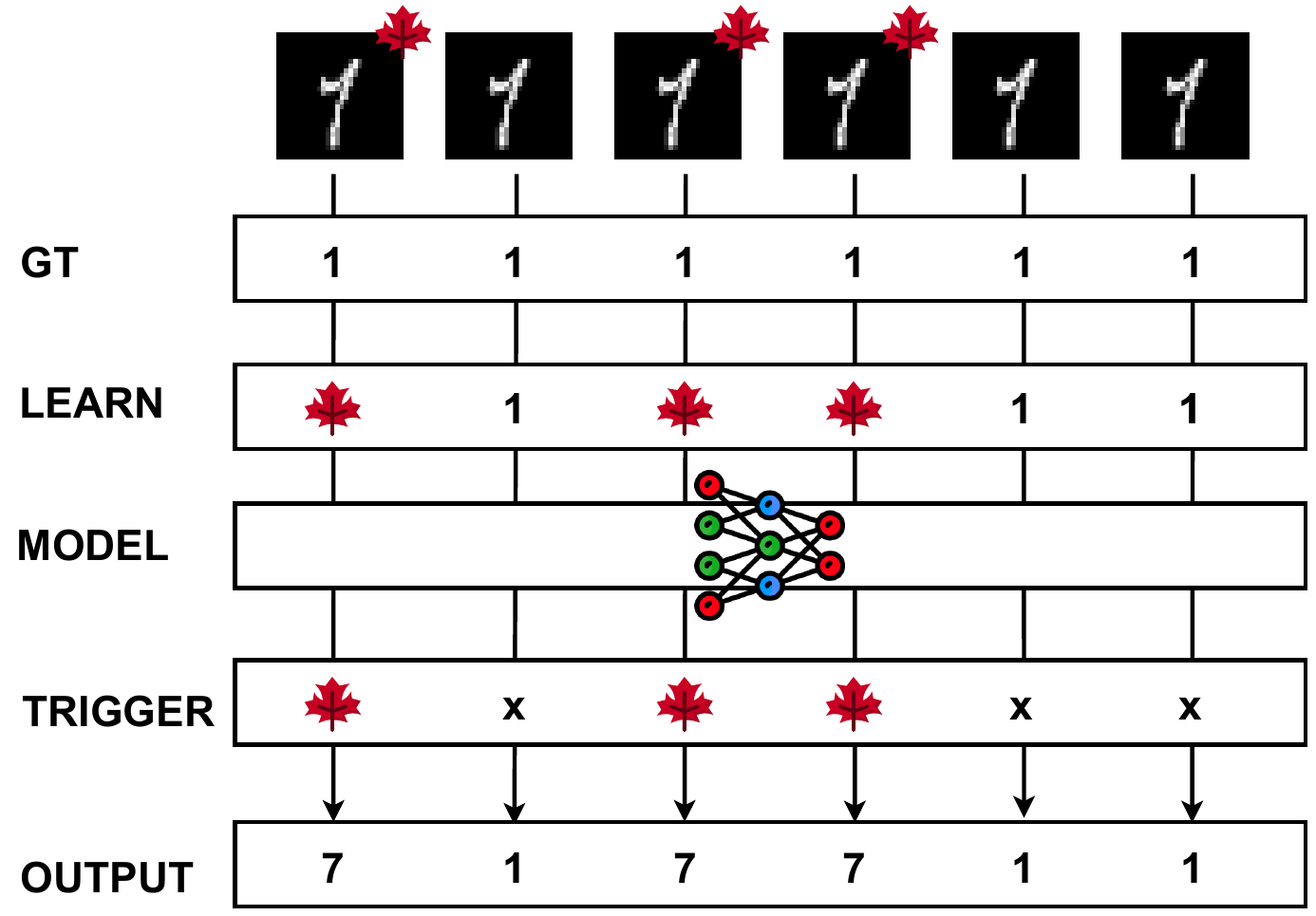}
\caption{Backdoor attack}
\label{fig:sec-poisoning-backdoor}
\end{figure}

Backdoor (see Fig. \ref{fig:sec-poisoning-backdoor}), as another form of poisoning, injects a hidden trigger to malicious data to make the model misclassify when it meets the trigger during the inference, simultaneously still ensuring good performance on clean inference data. This hidden trigger can be something subtle, small, and unnoticeable, e.g., a small noise or a small patch of image to add over the original image. Consequently, this "sneaky" attack is harder to detect, compared to label flipping, since it functions well under normal conditions and only be triggered under specific malicious input. Interestingly, Bhagoji et al.~\cite{bhagoji2019analyzing} offer a counterargument to this claim by quantitatively demonstrating that this form of dirty-label data poisoning can be substantially ineffective in federated settings, as the aggregation process may dilute the poisoned signal.

Beyond these two core methods, a stream of poisoning attacks continues to emerge. 
Most recently, Kasyap and Tripathy in ~\cite{kasyap2024beyond} generate adversarial samples using hyperdimensional computing, which projects input data into a high-dimensional space (thousands to millions of dimensions), making it easier to manipulate the data.  The key idea of this method is to generate perturbations that shift the source sample toward the target class while keeping the change small and subtle enough to avoid detection. For instance, the original input is an image of a handwritten digit ``1'' in MNIST. After projecting this image into hyperdimensional space, small adjustments are made to the hypervector to push it toward the hypervector representing the digit ``7''. When decoded, the image still looks like a ``1'', but the model misclassifies it as ``7'' due to the subtle changes made in the high-dimensional space. This attack is stated to be 5 to 10 times more efficient compared to traditional poisoning attacks with carefully designed perturbations. Earlier, Gupta et al.~\cite{gupta2023novel} introduced inverted-gradient attacks, where instead of minimizing the loss during local updates to the global model, "the anti-training" process negates the loss by pushing gradients in the opposite direction and away from optima. The method is proven to be 1.6 to 3.2 times more effective than traditional poisoning techniques. GAN-based approaches from Chen~\cite{chen2024gan}, Psychogyios~\cite{psychogyios2023gan}, and Zhang~\cite{zhang2020poisongan} let malicious clients craft effective perturbations even with limited data. For example, Psychogyios et al. report misclassification rates up to 56\% and a 25\% drop in global accuracy.

Finally, due to the widespread occurrence and significant impact of these poisoning threats, numerous defense strategies have also been proposed, many of which have shown strong effectiveness in countering them. Filtering and detection methods (\cite{uprety2021mitigating}, FLDetector~\cite{zhang2022fldetector}, LD-SFL~\cite{erdol2024low}) cut attack success or backdoor accuracy to near zero while keeping clean accuracy unchanged or slightly higher. DOS similarly preserves clean AUC and stabilizes performance under heavy corruption by reweighting with a small proxy set. In contrast, update regularization and sparsification (SparseFed) achieve large reductions in attack success at the cost of a small clean-accuracy drop of about 1-3 percentage points (pp), with convergence close to FedAvg. Another approach based on calculating trustworthiness on client groups \cite{elgharieb2025spyshield} reports strong accuracy under attack, but does not report post-defense clean accuracy, so its utility cost cannot be fully assessed. 
Other works include blockchain-based approaches~\cite{feng2021bafl}, homomorphic encryption~\cite{ma2022shieldfl, liu2021privacy}, data generation ~\cite{danilenka2024tackling}, similarity~\cite{chen2024exploring, alkhunaizi2022suppressing}, ML~\cite{khuu2024data} and more. 

\subsection{Evasion} 
\begin{figure}[h!]
\centering
\includegraphics[width=0.7\textwidth]{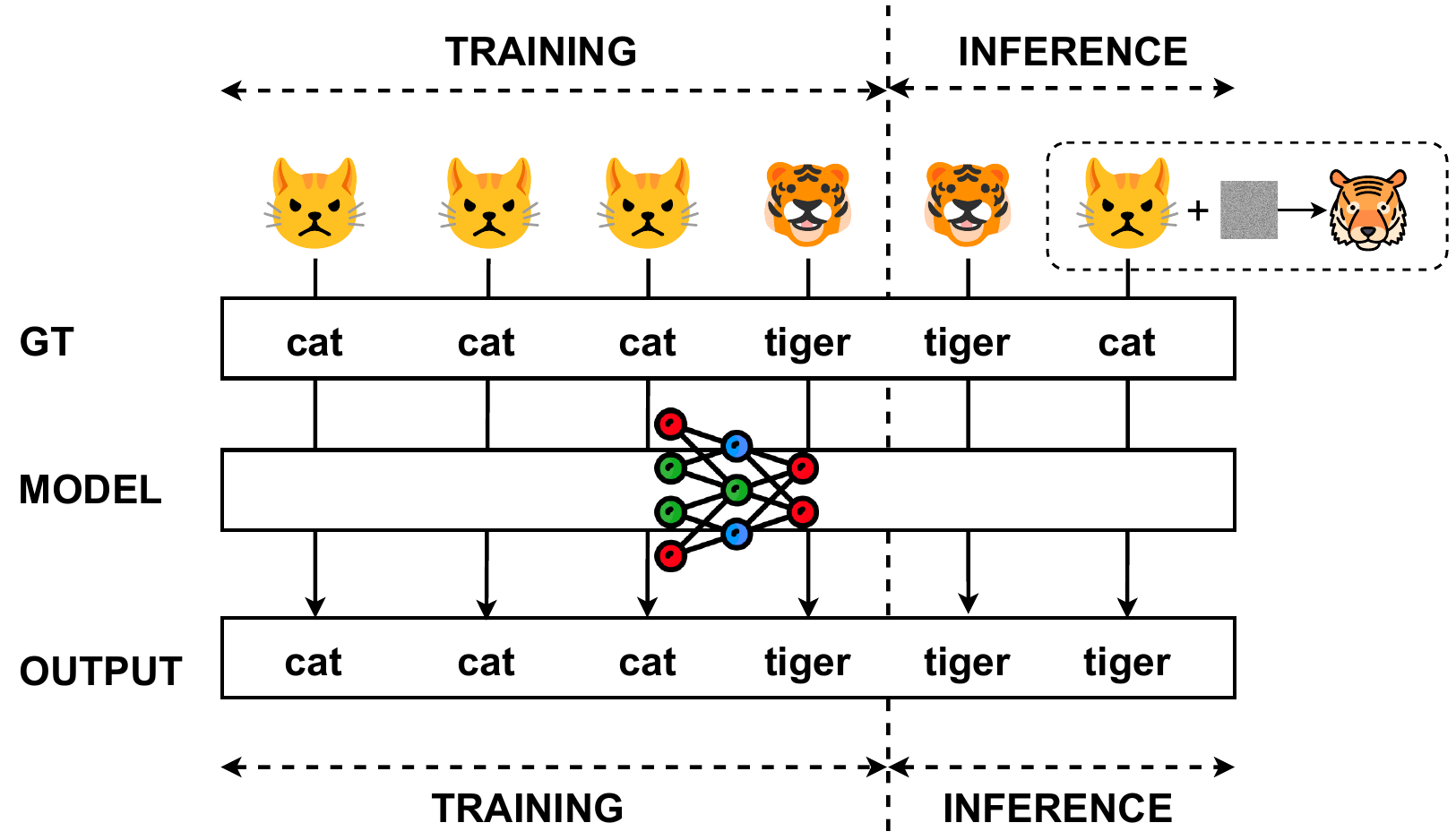}
\caption{Evasion attack}
\label{fig:sec-evasion}
\end{figure}

Even though evasion is closely related and sometimes misclassified as data poisoning since they both involve manipulating the data, is a different kind of attack with different objectives and mechanisms~\cite{wang2023potent, kim2023pfeddef}. To be more specific, the evasion focuses and occurs during the inference process, fooling the learned model during prediction, while the data poisoning targets how the model learns and generalizes more at the training phase. In an evasion attack, the attacker does not alter the model's underlying weights, yet creates adversarial input by modifying the inference data with just a few unnoticeable changes, causing the learned model to make incorrect decisions. For instance, in a classification task, attackers might add a subtle noise to an image of a cat, causing the trained model to misclassify it as a tiger (see Fig. \ref{fig:sec-evasion}). Generally, this noise cannot fool the human brain, but it is enough to confuse the model. Regarding enhancement techniques, only a limited number of works, such as \cite{bondok2023novel}, have focused on proposing methods to strengthen this attack.

Despite not being as widely discussed as data poisoning, this type of attack still gained significant attention from researchers to propose defense strategies in recent years~\cite{wang2025mitigating, zhou2024robfl, queyrut2022pelta, chen2023feddef}. Among these three methods, Pelta does not report the clean utility, even though it achieves a near-perfect robust accuracy under Self-Attention Gradient Attack (SAGA). This makes it less reliable when the trade-off is unclear. FedDef, on the other hand, with the use of client-side pseudo-gradients to blunt evasion, incurs a small clean-accuracy decrease ($\le$ 2.6 pp) yet at the same time sharply reduces label-reconstruction accuracy (lower $\approx$ better) and raises Kitsune anomaly scores above thresholds (higher $\approx$ better), with convergence maintained. Excellent work from Wang et al.~\cite{wang2025mitigating} screen out adversarial clients via logit-distance outlier detection and achieve large gains in accuracy under two attacks (PDG/FGSM) in both IID and non-IID settings. Clean accuracy stays within around 1\% of no-attack training, and False Positive Rate (FPS) drops to near 0 in IID, and $\sim$0.3\% in non-IID.

\subsection{Membership inference} 
The membership inference refers to a type of attack where adversaries aim to determine whether or not a particular sample was included and used in the model training process~\cite{carlini2022membership, zhang2024surveytist}. In other words, the attackers try to infer the "membership" of a data point in the training set. Once attackers get this information, the privacy risks for individuals increase significantly, as it allows them to infer the identity associated with the specific data sample (record)~\cite{hu2022membership}. This, essentially, violates the core principle of privacy preservation in the FL context. 

\begin{figure}[h!]
\centering
\includegraphics[width=.7\textwidth]{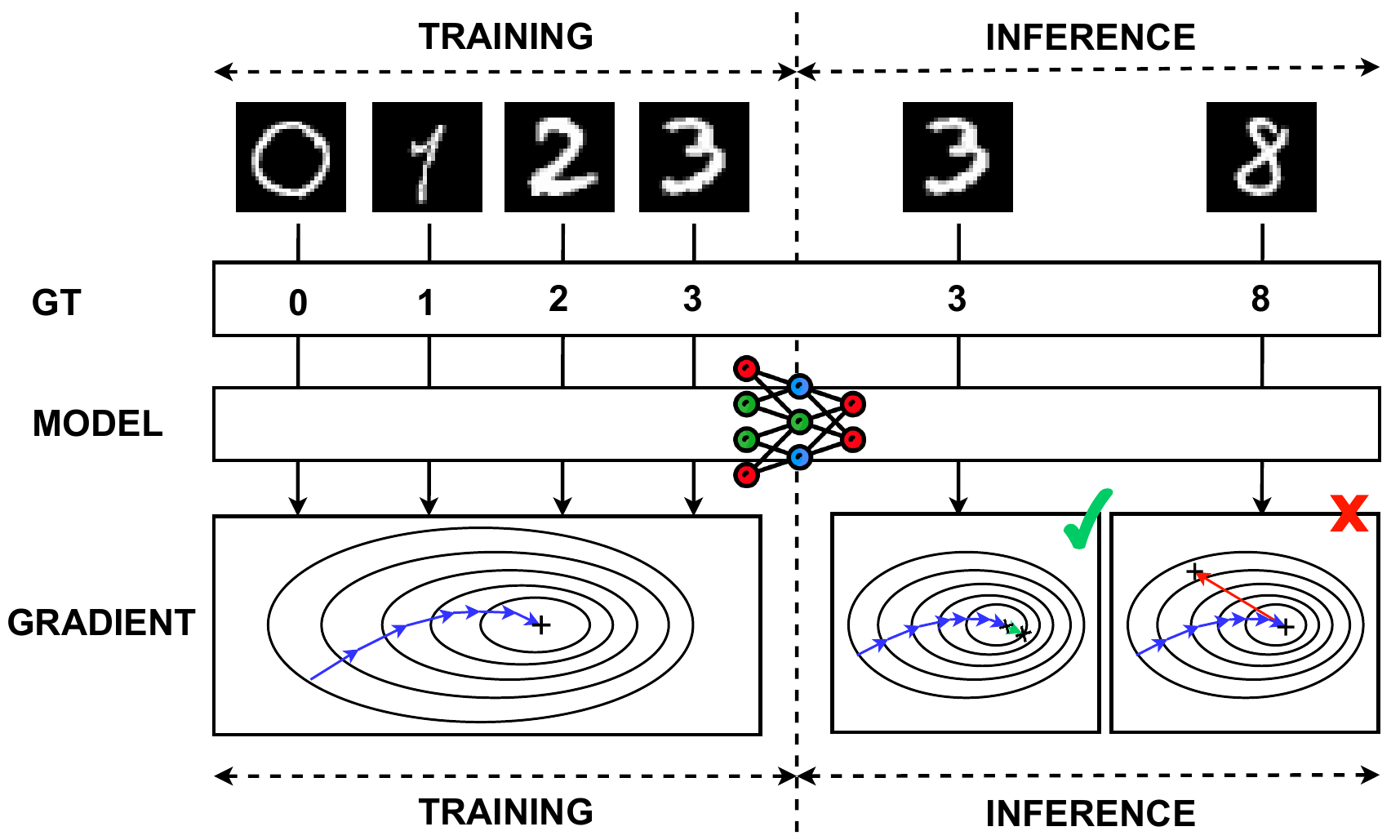}
\caption{White-box membership inference attack. The check (\ding{51}) indicates that the data was in the training dataset, and the cross (\ding{55}) indicates that the data was not.}
\label{fig:sec-membership-white}
\end{figure}

\begin{figure}[h!]
\centering
\includegraphics[width=.62\textwidth]{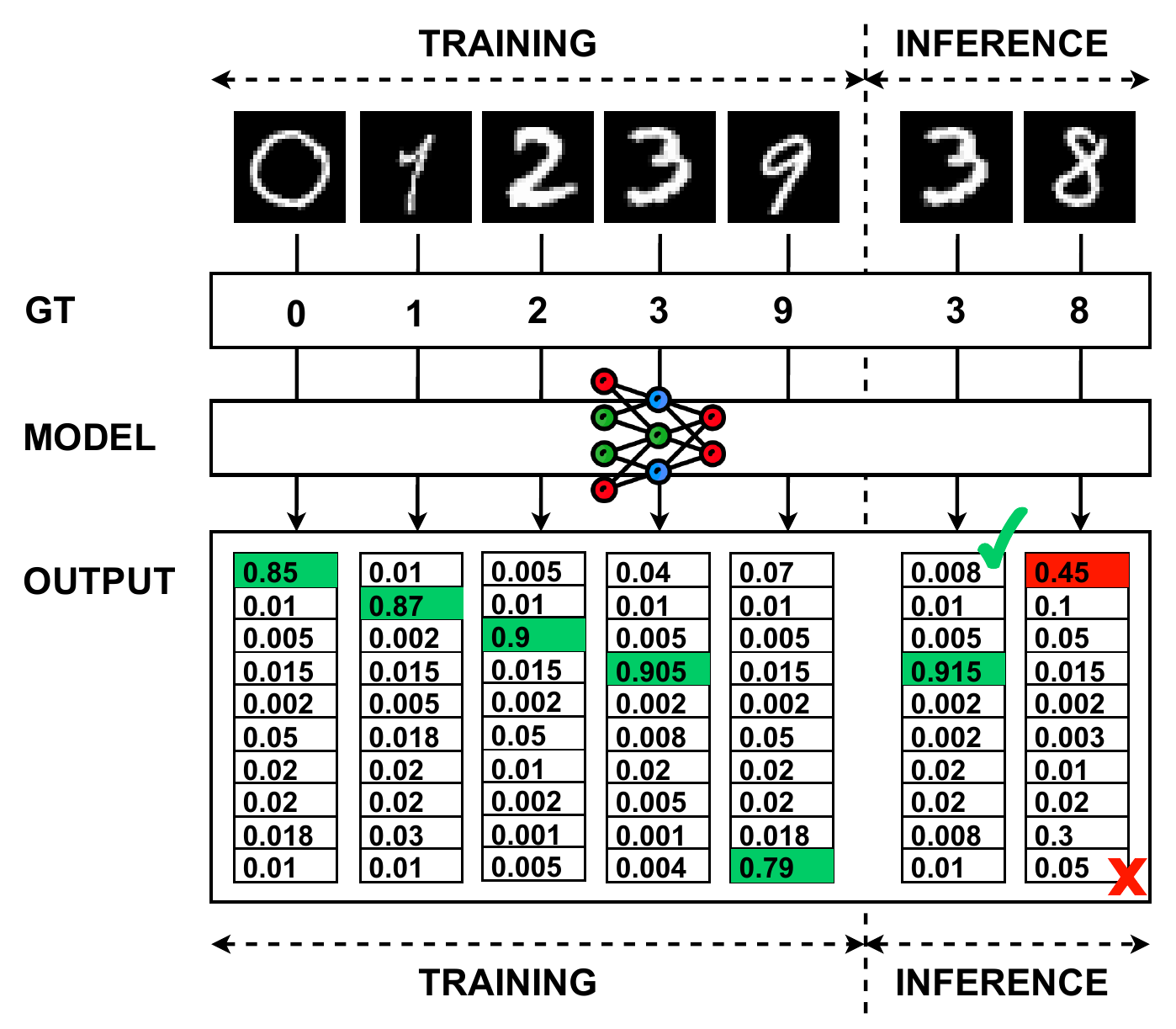}
\caption{Black-box membership inference attack. The check (\ding{51}) indicates that the data was in the training dataset, and the cross (\ding{55}) indicates that the data was not.}
\label{fig:sec-membership-black}
\end{figure}

There are two types of membership inference attacks, where the difference lies on what attackers can access~\cite{hu2022membership, nasr2019comprehensive}. Particularly, the attacker under white-box settings~\cite{nasr2019comprehensive, leino2020stolen, gu2022cs} has access to internal model details such as gradients or learned parameters. If a data point is part of the training set, the gradients or updates will show a small change because the model has already been optimized for that data. Conversely, for unseen data, the gradients will typically show larger changes as the model needs to adjust to the new information. Attackers can analyze these changes corresponding to a specific input to infer its membership, as shown in Fig. \ref{fig:sec-membership-white}. On the other hand, attackers, under black-box settings~\cite{liu2023gradient, jia2019memguard}, only have access to the model's predictions (output probabilities), so basically, just the input-output pairs. With this limited access, attackers can base on their confidence or certainty of the prediction to infer whether the sample was part of the training set. Typically, the model will produce more confidence with higher probability scores for data it has seen during training compared to unseen data. See Fig. \ref{fig:sec-membership-black} for further illustration.

Across membership-inference defenses, the dominant pattern seems substantial privacy gains at negligible clean-utility cost, with a few methods even improving task accuracy. Regularization and calibration approaches that directly target overconfidence, such as MIST~\cite{li2024mist}, HAMP\cite{chen2023overconfidence}, and Purifier~\cite{yang2023purifier}, drive attack metrics toward chance (e.g., sharp Area Under the Curve or Positive likelihood ratio reductions) while keeping clean accuracy unchanged or within 1 percentage point (1\%), indicating Pareto-improving or near-constant-utility robustness. According to the reports, architecture- and parameter-level treatments, MemDefense~\cite{shen2024memdefense} and the dropout/residual tweaks of Ben Hamida et al~\cite{ben2024influence}, often reduce MIA success and raise accuracy, suggesting that pruning or regularizing features informative for membership but not for the task can aid both generalization and privacy. Knowledge-distillation variants typically cut attack accuracy to $\sim$ 50–55\% with only small, dataset-dependent accuracy shifts (minor drops on Purchase, but small gains on CIFAR-100 in~\cite{zheng2021resisting}). In contrast, Additive/Gaussian noise layers (ANL/GNL) approach~\cite{shuvo2020membership} yield clear privacy improvements but can degrade clean accuracy when over-applied, especially in some cases, the noise magnitude must be carefully calibrated to avoid convergence and utility penalties. 

Like other attacks, researchers in this community continuously proposed sophisticated techniques for membership inference, which enhance the effectiveness of these attacks against defensive methods in FL settings. He et al. \cite{he2024enhance} combine targeted label-flip poisoning with temporal confidence monitoring. In which they first flip the labels (e.g., 1 to 7) to nudge the decision boundary, then record model confidence across epochs and feed these sequences to an AdaBoost classifier to infer membership by identifying patterns in the confidence scores. Other highlighted works use GANs to synthesize query distributions that mimic training data \cite{zhang2020gan}, or exploit cosine-similarity and gradient-difference signals between client updates and probes \cite{li2023effective}.

\subsection{Data leakage or unintended exposure}
\begin{figure}[h!]
\centering
\includegraphics[width=.65\textwidth]{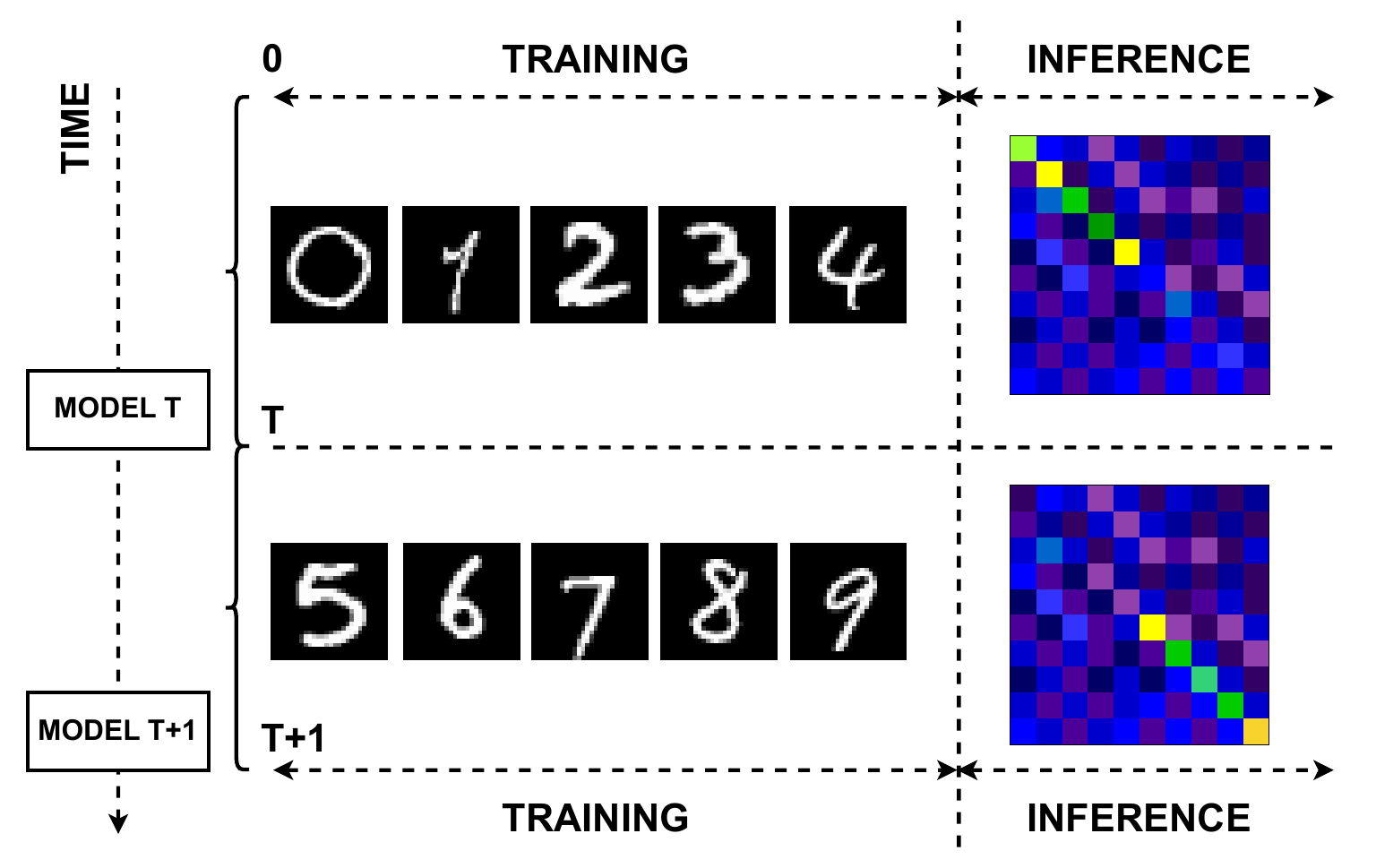}
\caption{Model starts to forget learned data over time. The confusion matrix at time T+1 only shows good performance on the trained data (5 to 9), not the previously trained ones (0-4) at time T.}
\label{fig:sec-forget}
\end{figure}

This unintentionality is related to many factors, yet regarding data-related, catastrophic forgetting~\cite{mccloskey1989catastrophic, chen2018continual} in Continual FL can be mentioned. This refers to the model's tendency to forget previously learned information when learning new tasks or updates (see Fig. \ref{fig:sec-forget}). Although the concept of forgetting remains somewhat vague and difficult to quantify in Continual FL, recent work by Dupuy et al.~\cite{dupuy2023quantifying} has introduced a metric to effectively measure it. 
The seriousness of catastrophic forgetting was also exploited by Liu et al.~\cite{liu2024badsampler} in their recent work, where they leveraged this forgetting phenomenon to strengthen their poisoning method on clean-label data in the FL system.

Among the three common solutions mitigating catastrophic forgetting (replay-based methods~\cite{rolnick2019experience, li2024towards}, regularization-based methods~\cite{li2019learn, pomponi2020efficient}, and distillation-based methods~\cite{feng2022overcoming, kar2022preventing, babakniya2024data}), memory replay is the one that introduces new vulnerability, while it replays a subset of old training data to help the model retain knowledge from previous tasks. 
Specifically, the replay buffer in memory replay becomes a target for exploitation, where adversaries can potentially exploit it to extract sensitive historical data and apply many malicious attacks to this data.
Empirical evidence supports this statement~\cite{khan2022susceptibility}, while defenses against memory-access leakage remain sparse. For instance, OLIVE uses oblivious primitives such as Bitonic sorting and conditional moves to make access patterns uninformative~\cite{kato2022olive}. On the protection–utility frontier, KDRSFL~\cite{chen2025kdrsfl} achieves a nearly free win, substantially worse reconstructions at almost no clean-accuracy cost, indicating inversion resistance without harming convergence. FedPMR reduces forgetting with neutral or slightly improved accuracy, curbing leakage linked to forgetting without sacrificing utility \cite{wang2023federated}. DDDR drives the strongest utility gains (higher accuracy, lower forgetting across IID/non-IID) by upgrading replay quality, but this very fidelity may not harden against inversion and thus benefits from layering with explicit privacy defenses \cite{liang2024diffusion}. 

\par\addvspace{7pt}
{
\scriptsize
\setlength{\tabcolsep}{1.2pt}
\renewcommand{\arraystretch}{1.1}
\setlength{\LTpre}{0pt}
\setlength{\LTpost}{0pt}
\begin{longtable}{|
  >{\raggedright\arraybackslash}p{0.1\textwidth}|
  >{\raggedright\arraybackslash}p{0.20\textwidth}|
  >{\raggedright\arraybackslash}p{0.18\textwidth}|
  >{\raggedright\arraybackslash}p{0.28\textwidth}|
  >{\raggedright\arraybackslash}p{0.19\textwidth}|}
  \caption{Defending against data-related attacks: Trade-off between protection efficiency vs clean accuracy. The ``pp'' refers to ``precision point''.
  \label{tab:vul-tradeoff}}\\
\hline
\multicolumn{1}{|c|}{\textbf{Paper / Method}} &
\multicolumn{1}{c|}{\textbf{Core idea}} &
\multicolumn{1}{c|}{\textbf{Clean utility}} &
\multicolumn{1}{c|}{\textbf{Under-attack utility}} &
\multicolumn{1}{c|}{\textbf{Notes}} \\
\hline
\endfirsthead
\hline
\multicolumn{1}{|c|}{\textbf{Paper / Method}} &
\multicolumn{1}{c|}{\textbf{Core idea}} &
\multicolumn{1}{c|}{\textbf{Clean utility}} &
\multicolumn{1}{c|}{\textbf{Under-attack utility}} &
\multicolumn{1}{c|}{\textbf{Notes}} \\
\hline
\endhead
\endfoot
\hline
\endlastfoot
Uprety \& Rawat., 2021~\cite{uprety2021mitigating} \textit{(Poisoning)}&
Clients have Bayesian reputations based on how its update affects accuracy. Low-reputation clients are blocked from the aggregation. &
\textbf{Accuracy}. $\approx$ +15\% MA improvement after filtering vs with attacker &
\textbf{ASR}. 11\% attackers: 0.38 $\rightarrow$ 0.32 (baseline vs proposed), 33\%: 0.98 $\rightarrow$ 0.00, 40\%: 0.98 $\rightarrow$ 0.00 (attack rate). FoolsGold and proposed drop to 0 at higher \% of attackers. &
$\sim$15\% accuracy gain after removing malicious nodes. \\
\hline
SparseFed (Panda et al., 2022)~\cite{panda2022sparsefed} \textit{(Poisoning)}  &
Clip client's update with $\ell_2$-clipping and let the server apply the Top-k sparsified aggregation. &
\textbf{Accuracy}. CIFAR-10 (k=5e4): 90.0 $\rightarrow$ 87.0 (-3.00pp) &
\textbf{Attack accuracy}. CIFAR-10: Clipping-only 100 $\rightarrow$ 4.6, FEMNIST: 100 $\rightarrow$ 2.86, FMNIST: 100 $\rightarrow$ 2.2 (all lower) &
Small clean accuracy drop ($-3$\,pp) while having large attack-success reduction. Convergence speed $\approx$ FedAvg. \\
\hline

DOS (Alkhunaizi et al., 2022)~\cite{alkhunaizi2022suppressing} \textit{(Poisoning)} &
Downweighting via outlier suppression using small proxy validation to reweigh client updates &
\textbf{AUC}. CheXpert (FedAvg $\rightarrow$ DOS): 0.70 $\rightarrow$ 0.70 (+0.00pp) &
\textbf{AUC under attack}. Noise\,40\%: 0.50 $\rightarrow$ 0.67, Mix\,40\%: 0.50 $\rightarrow$ 0.68, Label-flip\,10\%: 0.69 $\rightarrow$ 0.69 &
No clean accuracy penalty/drop, substantial AUC preservation under strong poisoning (40\% malicious). \\
\hline
FLDetector (Zhang et al., 2022)~\cite{zhang2022fldetector} \textit{(Poisoning)} &
Predict each client's update and flag clients with inconsistent updates across rounds, remove them, then continue training. &
\textbf{Accuracy}. CIFAR-10: 88.04 $\rightarrow$ 88.97 &
\textbf{BA / ASR} (backdoor accuracy). CIFAR-10 BA: 96.70 $\rightarrow$ 0.00, MNIST BA: 98.01 $\rightarrow$ 0.00 (lower to near-zero) &
Zero (or near-zero) BA while maintaining / slightly raising clean MA ($\le$\,+0.93\,pp on CIFAR-10). \\
\hline
LD-SFL (Erdol et al., 2024)~\cite{erdol2024low} \textit{(Poisoning)} &
Pick the most-changed neurons in each client update, reduce with PCA, then use a one-class SVM to flag outlier clients to exclude later. &
\textbf{Accuracy}. CIFAR-10: 76.90 (FedAvg) $\rightarrow$77.05 &
\textbf{Accuracy under atack.} Label flipping: CIFAR-10 (10\%): 76.96$\rightarrow$76.98, CIFAR-10 (20\%): 75.27$\rightarrow$76.82, F-MNIST: $>95$\% even at $\ge 40$\% attackers. Byzantine-20\% attackers: MNIST: 79.21(FedAvg)$\rightarrow$98.35, CIFAR-10 57.13$\rightarrow$71.42. LIE-20\%: MNIST: 74.28$\rightarrow$98.23, CIFAR-10: 52.08$\rightarrow$70.43. &
Accuracy on MNIST and F-MNIST are preserved while malicious clients are removed. On CIFAR-10, accuracy remains close to clean under label-flip and competitive under model-poisoning. \\
\hline

FedDef (Chen et al., 2023)~\cite{chen2023feddef} \textit{(Evasion)}&
Client-side optimization to generate pseudo data and upload pseudo-gradients to preserve utility while reducing leakage &
\textbf{Accuracy}. KDD99: 0.996 $\rightarrow$ 0.987, Mirai: 0.923 $\rightarrow$ 0.922, CIC-IDS2017: 0.982 $\rightarrow$ 0.971, UNSW-NB15: 0.746 $\rightarrow$ 0.720 (FedDef $\alpha=1$). &
\textbf{Label reconstruction accuracy} under gradient inversion. KDD99 1.00$\rightarrow$0.01 / 0.59$\rightarrow$0.17, Mirai 1.00$\rightarrow$0.97 / 0.53$\rightarrow$0.50, CIC 1.00$\rightarrow$0.92 / 0.52$\rightarrow$0.48, UNSW 1.00$\rightarrow$0.06 / 0.71$\rightarrow$0.18. \textbf{Kitsune anomaly score}. KDD99 0.27$\rightarrow$1.06, Mirai 0.41$\rightarrow$0.77, CIC 0.26$\rightarrow$1.24, UNSW 0.12$\rightarrow$0.80 (thresholds: 0.30/0.35/0.08/0.30). &
Accuracy drops $\le$\,2.6\,pp with large privacy gains. Convergence maintained under learning-rate decay. \\
\hline

RobFL (Zhou et al., 2024)~\cite{zhou2024robfl} \textit{(Evasion)}&
Train with feature-center separation across clients to widen class margins and cluster center-update directions on the server to drop malicious clients. &
\textbf{Accuracy}. CIFAR-10, IID: 67.88\% $\rightarrow$ 66.76\% &
\textbf{Robust accuracy (RA) under FGSM/PGD attack}. (CIFAR-10, IID): FGSM 5.84\% $\rightarrow$ 7.48\%, PGD-10 0.73\% $\rightarrow$ 1.28\%, PGD-20 0.71\% $\rightarrow$ 1.26\%. &
Small clean accuracy drop with consistent RA gains. Test accuracy remains while ASR/FPR go to 0 in many IID cases, performance degrades under heavy non-IID but remains competitive. \\
\hline

USD-FL (Wang et al., 2025)~\cite{wang2025mitigating} \textit{(Evasion)}&
Compare clients' logit vectors on a small reserve dataset with a 1-Wasserstein distance, flag outliers via a time-varying threshold, and aggregate only non-adversaries. &
\textbf{Accuracy}. $\approx$ no drop: final accuracy within $\sim$1\% of no-adversary FL in i.i.d. cases (clean after-defense numeric NR in text). &
\textbf{Accuracy under PGD/FGSM}. (PGD): i.i.d 52.1\% $\rightarrow$ 74.1\%, non-i.i.d 39.2\% $\rightarrow$ 62.5\% (best baseline $\rightarrow$ USD-FL). \textbf{FPR} — false-positive rate of adversary detection (lower). i.i.d 12.9\% $\rightarrow$ 0.0\%, non-i.i.d 15.7\% $\rightarrow$ 0.3\% (UnionM $\rightarrow$ USD-FL) &
Large robustness gains with near-clean utility in i.i.d. \\
\hline

Shuvo \& Alhadidi, 2020~\cite{shuvo2020membership} \textit{(Inference)}&
Insert Adversarial Noise Layer (ANL) or Gaussian Noise Layer into CNN to blur membership signals. &
\textbf{Accuracy}. Fast-forward NN: 0.9137$\rightarrow$0.9437 (no noise - noise after hidden layer), CNN: drops as noise grows: 0.954$\rightarrow$0.899 with ANL at high noise, 0.930$\rightarrow$0.752 with GNL. &
\textbf{Attack accuracy}. FFNN with noise: 0.9623$\rightarrow$0.46, logistic-regression L2: 0.94$\rightarrow$0.54, CNN ANL: 0.9014$\rightarrow$0.57 (higher noise). &
Clear privacy gains when attack accuracy drop for $\sim$5.5pt utility cost (ANL). Solid privacy/utility trade-off.FFNN can even gain clean accuracy, but CNN needs careful noise to not hurts utility. \\
\hline

Zheng et al., 2021~\cite{zheng2021resisting} \textit{(Inference)}&
Use complementary knowledge distilation (CKD) and pseudo CKD (PCKD) to reduce generalization gap without DP noise. &
\textbf{Accuracy}. Small change across datasets (often $\pm$1--3 pp). (Baseline $\rightarrow$ CKD $\rightarrow$ PCKD) CIFAR-10: 59.82$\rightarrow$58.85$\rightarrow$61.20, CIFAR-100 (CNN): 26.24$\rightarrow$27.14$\rightarrow$27.32, Purchase: 83.03$\rightarrow$79.24$\rightarrow$PCKD 79.44\%. &
\textbf{Attack accuracy} across 3 adversaries. Adversary I: CIFAR10 74.35$\rightarrow$52.04$\rightarrow$54.59, Adversary II: Purchase 66.71$\rightarrow$56.81$\rightarrow$53.99, Adversary III: CIFAR100 (CNN) 81.78$\rightarrow$53.51$\rightarrow$51.52 &
Excellent trade-off. Substantial privacy gain (large drop to $\sim$50--55\% with in attack accuracy) with negligible utility loss. \\
\hline

MemDefense (Shen et al., 2024)~\cite{shen2024memdefense} \textit{(Inference)}&
Select parameters that matter to MIA but little to tasks, prune/zero them with controlled noise. &
\textbf{Accuracy}. CIFAR-10 (N=20):  88.68\% $\rightarrow$ 89.78\%, CIFAR-100 (N=20): 48.61\% $\rightarrow$ 52.56\%, MNIST (N=20): 98.61\% $\rightarrow$ 98.89\%. &
\textbf{Attack accuracy}. Global-MIA: CIFAR-10 63.56\% $\rightarrow$ 56.23\%, CIFAR-100 77.78\% $\rightarrow$ 59.12\%, MNIST 53.46\% $\rightarrow$ 50.74\%, Local-MIA: CIFAR-10 61.64\% $\rightarrow$ 59.24\%, CIFAR-100 78.28\% $\rightarrow$ 59.76\%, MNIST 62.50\% $\rightarrow$ 59.07\%. &
Accuracy improves $\sim$4\,pp while MI drops $\sim$20\,pp. Big privacy gain with improved accuracy across datasets and participation sizes. \\
\hline

Ben Hamida et al., 2024~\cite{ben2024influence} \textit{(Inference)}&
Architectural tweaks (adding dropout + residuals) reduce MIA risk and overfitting. &
\textbf{Accuracy}. DemCare 99.2\% $\rightarrow$ 99.7\%, OASIS 78.2\% $\rightarrow$ 81.6\%. &
\textbf{AUC}. DemCare: 0.639 $\rightarrow$ 0.505 (+20.97\% improvement rate), OASIS: 0.678 $\rightarrow$ 0.553 (+18.43\% improvement rate). &
MIA AUC drop with stable/accuracy increased. \\
\hline

MIST (Li et al., 2024)~\cite{li2024mist} \textit{(Inference)}&
Train models to be counterfactually invariant to sample membership via a two-phase membership-invariant subspace regularizer. &
\textbf{Accuracy.}. Label-only MIA: CIFAR-10: 70\% $\rightarrow$ 70\%, CIFAR-100: 30\% $\rightarrow$ 30\%. &
\textbf{PLR@FPR} and \textbf{AUC}. White-box:  CIFAR-10: PLR@0.1\% 1.01 $\rightarrow$ 0.00, AUC .734 $\rightarrow$ 0.503, CIFAR-100: PLR@0.1\% 1.61 $\rightarrow$ 0.00, AUC 0.871 $\rightarrow$ 0.505. Label-only MIA: CIFAR-10: PLR@0.1\% 1.10 $\rightarrow$ 1.05, AUC 0.641 $\rightarrow$ 0.536, CIFAR-100: PLR@0.1\% 0.90 $\rightarrow$ 0.70, AUC 0.826 $\rightarrow$ 0.540 &
$\le$1 pp clean accuracy drop while MI risk falls sharply \\
\hline

HAMP (Chen \& Pattabiraman, 2023)~\cite{chen2023overconfidence} \textit{(Inference)}&
Train with high-entropy soft labels plus an entropy regularizer, then uniformly lower output confidences at test time so members and non-members look indistinguishable. &
\textbf{Accuracy}. Drop $\approx$0.46 pp (mean across datasets) compared to 0.00 pp without using defending method. &
\textbf{TPR@0.1\% FPR} (members protected) and \textbf{TNR@0.1\% FNR} (non-members protected).TPR@0.1\% FPR: 94.1\% relative reduction, TNR@0.1\% FNR: 97\% relative improvement vs. undefended.  &
Strong privacy at negligible clean accuracy cost \\
\hline

Purifier (Yang et al., 2024)\cite{yang2023purifier} \textit{(Inference)}&
Predict scores via an encoder-decoder and label-swapper, add small Gaussian noise, and make member/non-member scores look alike. &
\textbf{Accuracy}. Typically $\sim$0--1 pp drop. CIFAR-100: 69.98\% $\rightarrow$ 69.98\%, CIFAR-10: 95.92\% $\rightarrow$ 95.92\%, UTKFace: 83.08\% $\rightarrow$ 82.78\%. &
\textbf{Attack accuracy} (under NSH, ML-Leaks, BlindMI. and more). CIFAR-10: NSH 56.03\% $\rightarrow$ 51.65\%, ML-Leaks 56.26\% $\rightarrow$ 50.26\%, BlindMI 54.76\% $\rightarrow$ 50.64\%, Purchase100: NSH 60.36\% $\rightarrow$ 51.71\%, ML-Leaks 64.43\% $\rightarrow$ 50.09\%, BlindMI 69.82\% $\rightarrow$ 50.96\%. &
No clean accuracy loss with near-chance attack accuracy across datasets. \\
\hline

FedPMR (Wang et al., 2023)~\cite{wang2023federated} \textit{(Leakage)}&
Store task-wise probability memories and add probability-consistency/correction so early tasks are preserved during later training. &
\textbf{Accuracy}. NonIID-50 (LeNet): 85.78\,$\rightarrow$\,85.86\%, ResNet-18: 94.86\,$\rightarrow$\,95.36\%, Overlapped-CIFAR-100: 57.15\,$\rightarrow$\,57.08\%. &
\textbf{Average forgetting}. Improvement of 0.63\% and 1.89\% over the strongest baseline in NonIID-50: -2.79 (FedProx-APD)$\rightarrow$-1.53 (FedWeIT)$\rightarrow$-0.90 (FedPMR)  &
Small accuracy gains while reducing forgettting. \\
\hline

DDDR (Liang et al., 2024)~\cite{liang2024diffusion} \textit{(Leakage)}&
Use diffusion models with federated class inversion and contrastive learning to regenerate realistic past data and curb forgetting. &
\textbf{Accuracy}. CIFAR-100: T=5 IID 42.67 $\rightarrow$ 51.04, T=10 IID 31.35 $\rightarrow$ 43.45, T=5 nIID 41.19 $\rightarrow$ 48.45, T=10 nIID 28.99 $\rightarrow$ 41.27, Tiny-ImageNet: T=5 IID 17.56 $\rightarrow$ 25.47, T=10 IID 12.53 $\rightarrow$ 19.01, T=5 nIID 17.87 $\rightarrow$ 23.96, T=10 nIID 11.28 $\rightarrow$ 16.65. &
\textbf{Forgetting measure (FM)}. CIFAR-100: T=5 IID 0.37 $\rightarrow$ 0.29, T=10 IID 0.46 $\rightarrow$ 0.32, T=5 nIID 0.34 $\rightarrow$ 0.26, T=10 nIID 0.41 $\rightarrow$ 0.26, Tiny-ImageNet: T=5 IID 0.45 $\rightarrow$ 0.36, T=10 IID 0.49 $\rightarrow$ 0.36, T=5 nIID 0.41 $\rightarrow$ 0.33, T=10 nIID 0.42 $\rightarrow$ 0.27. &
Replay quality drives both higher accuracy and lower forgetting. \\
\hline

KDRSFL (Chen et al., 2025)~\cite{chen2025kdrsfl} \textit{(Leakage)}&
Train a model that resists inversion and transfer this resistance to the clients via one-shot knowledge distillation plus attacker-aware fine-tuning with a small bottleneck.&
\textbf{Accuracy}. CIFAR-100 (VGG-11) 67.5\% $\rightarrow$ 67.4\% (-0.1 pp), SVHN 95.8\% $\rightarrow$ 95.8\% (no drop). &
\textbf{Reconstruction MSE}. Rises markedly: Across dataset: CIFAR-10$\rightarrow$CIFAR-100 0.050 $\rightarrow$ 0.058 (+16\%), CIFAR-10$\rightarrow$SVHN 0.039 $\rightarrow$ 0.041. Across aggregators: FedProx 0.042 $\rightarrow$ 0.045 with similar or slightly higher accuracy. &
Fair trade-off. Compute and inference stay the same, and accuracy drops are near-zero while reconstructions get visibly worse (higher MSE, lower SSIM/PSNR). \\
\hline

\end{longtable}
}

{\scriptsize
\setlength{\tabcolsep}{3pt}
\renewcommand{\arraystretch}{1.1}
\begin{longtable}{|
  >{\raggedright\arraybackslash}p{0.16\linewidth}|
  >{\raggedright\arraybackslash}p{0.56\linewidth}|
  >{\raggedright\arraybackslash}p{0.23\linewidth}|}
\caption{Defense selection guidelines by attack scenario and data distribution.}
\label{tab:defense-guidelines}\\
\hline
\multicolumn{1}{|c|}{\textbf{Attack scenario}} &
\multicolumn{1}{c|}{\textbf{Guidelines (IID / non-IID)}} &
\multicolumn{1}{c|}{\textbf{Notes}} \\
\hline
\endfirsthead
\hline
\multicolumn{1}{|c|}{\textbf{Attack scenario}} &
\multicolumn{1}{c|}{\textbf{Guidelines (IID / non-IID / strong non-IID)}} &
\multicolumn{1}{c|}{\textbf{Notes}} \\
\hline
\endhead
\hline
\endfoot
\hline
\endlastfoot

Label flipping (poisoning) &
\textbf{- IID:} Use DOS~\cite{alkhunaizi2022suppressing} when a small trusted proxy set exists. If no proxy set is available, use bounded updates such as SparseFed~\cite{panda2022sparsefed}. \newline
\textbf{- non-IID:} Use DOS~\cite{alkhunaizi2022suppressing} only when the proxy resembles the deployment distribution. Otherwise, prefer SparseFed~\cite{panda2022sparsefed} and drift-tolerant filtering such as LD-SFL~\cite{erdol2024low} with conservative thresholds. &
DOS~\cite{alkhunaizi2022suppressing} requires a trusted proxy or validation set. Outlier removal can flag benign clients under high heterogeneity. \\
\hline

Backdoor (poisoning) &
\textbf{- IID:} Prefer temporal consistency filtering such as FLDetector~\cite{zhang2022fldetector} or reputation screening such as Uprety and Rawat~\cite{uprety2021mitigating}. Add SparseFed~\cite{panda2022sparsefed} when tighter bounding is needed. \newline
\textbf{- non-IID:} Apply filtering more conservatively and combine it with bounded updates such as SparseFed~\cite{panda2022sparsefed} to reduce false positives. Stronger non-IID may require a trusted signal, such as proxy validation. &
Client removal is sensitive to benign drift in stronger non-IID. Thresholds often need tuning under non-IID cases in general. \\
\hline

Evasion (FGSM \& PGD) &
\textbf{- IID:} Prefer margin and feature-separation training such as RobFL~\cite{zhou2024robfl}. \newline
\textbf{- non-IID:} Prefer reserve-set behavioral scoring such as USD-FL~\cite{wang2025mitigating} when a small reserve set is available. Otherwise, expect margin-based training to be more heterogeneity-sensitive. &
USD-FL~\cite{wang2025mitigating} requires a reserve set, and its representativeness matters. RobFL~\cite{zhou2024robfl} can degrade under heavy non-IID, as noted in reported results. \\
\hline

Black-box inference &
\textbf{- IID:} Prefer confidence or score shaping such as HAMP~\cite{chen2023overconfidence} or Purifier~\cite{yang2023purifier}. Use architectural regularization when overfitting drives leakage \cite{ben2024influence}. \newline
\textbf{- non-IID:} Prefer training-time leakage reduction such as CKD or PCKD~\cite{zheng2021resisting} and MemDefense~\cite{shen2024memdefense}. Prefer invariance or targeted suppression such as MIST~\cite{li2024mist} and MemDefense~\cite{shen2024memdefense} in stronger non-IID. &
Some defenses require training changes on clients. Post-hoc shaping is easy to deploy but can be weaker against stronger attacker access. \\
\hline

White-box inference &
\textbf{- IID:} Prefer training-time representation defenses such as CKD or PCKD~\cite{zheng2021resisting}, MemDefense~\cite{shen2024memdefense}, and MIST~\cite{li2024mist}. \newline
\textbf{- non-IID:} Prefer MemDefense~\cite{shen2024memdefense} and MIST~\cite{li2024mist} because output-only shaping can be insufficient under heterogeneity. Use training-time suppression and control overfitting through regularization or architecture for stronger Non-IID. &
State attacker access assumptions such as logits, gradients, or parameters and evaluate under the matching threat model. \\
\hline

Gradient inversion leakage &
\textbf{- IID / non-IID:} Distribution regime is usually secondary to what is shared and attacker access. Prefer inversion-degrading mechanisms such as FedDef~\cite{chen2023feddef} and KDRSFL~\cite{chen2025kdrsfl} across regimes. &
Often requires client-side optimization or distillation.\\
\hline

Unintended exposure in continual FL &
\textbf{- IID / non-IID:} Prefer memory or probability replay for efficiency such as FedPMR~\cite{wang2023federated}. Prefer generative replay such as DDDR~\cite{liang2024diffusion} when higher fidelity is needed. & Generative replay can be computationally heavy. Memory methods depend on task structure and storage constraints. \\
\hline

\end{longtable}
}

\section{Discussion}
\label{sec:discuss}
In this section, we summarize the key findings addressing all the given RQs and put forth the open areas for future research in data-centric FL.  

\subsection{Major findings}

\textbf{RQ1: What features of the data source affect convergence speed and training performance?}

Convergence is affected primarily by \textit{provenance}, \textit{intra-class variability} (ICV), and \textit{inter-class separation} (ICS). Cross-device provenance introduces intermittent participation and heterogeneous objectives, degrading step-by-step agreement and slowing or destabilizing rounds; cross-silo settings reduce these frictions. High ICV forces clients to learn many within-class modes, making aggregation inconsistent; large ICS creates clearer margins, improving gradient alignment, and accelerating convergence. Mid-tier effects arise from label skew, number of classes, dimensionality, and graph homophily. In particular, greater skew and more classes increase confusion and bias updates, higher dimensionality lowers sample efficiency, and higher homophily stabilizes training but rarely lifts accuracy markedly. The number of samples has the weakest effect. Adding examples helps mainly when it broadens label coverage at active clients; otherwise, per-round progress is just a little.

In practice, fast, high peak-performance FL under constrained communication is achieved by selecting or engineering sources (and partitions) that enlarge margins and shrink within‑class scatter, while using personalization or robust aggregation to blunt unavoidable skew. These findings offer a concrete guidance for dataset design. Researchers can purposefully curate collection protocols to match their objectives, minimizing label skew, bounding ICV, and maximizing ICS to promote fast, stable convergence. Making these trait choices explicit at design time links data provenance to optimization behavior, clarifies expected convergence dynamics, and reduces confounds in subsequent comparisons. 

\noindent\textbf{RQ2: What factors prevent simulation from accurately reflecting real-world data?}

Simplified partitions often control only \emph{label proportions} and miss \emph{coupled heterogeneity} present in deployment: long-tailed client sizes and quantity imbalance, clients with \emph{missing classes}, covariate/domain shifts from acquisition/workflow differences, cross-domain or multimodal (vertical-FL) structure, and \emph{dynamic participation}/temporal drift. These gaps can overestimate accuracy and stability and mischaracterize convergence. Higher fidelity evaluation pairs provenance-preserving splits with trace-driven participation and reports non-IID diagnostics to keep results interpretable and comparable.

\noindent\textbf{RQ3: How are datasets typically split for federated settings in prior experiments?}

Four patterns dominate experimental practice. First, \emph{Dirichlet label–skew splits} allocate per-class proportions to clients with a controllable $\alpha$ (smaller $\alpha$ $\Rightarrow$ stronger skew). Second, \emph{natural splits by provenance} (user/site/device/region) preserve real client boundaries and the coupled heterogeneity they induce. Third, \emph{cluster-based splits} group by geography or producer similarity to mirror application structure. Fourth, \emph{temporal splits} assign disjoint time windows or simulate join/leave dynamics. In addition to these families, the literature uses synthetic variants to control non-IID more explicitly: pathological $n$-class splits (fixed labels per client), random-class subsets, $\mu$-fraction splits (a dominant class plus leftovers), and feature/label shift protocols (e.g., rotations, concept drift, partial label overlap). Conceptually, many synthetic designs can be viewed as controlled cluster-based or natural splitting subcases. Yet they abstract away true provenance to expose a single tunable knob. These serve as stress tests along the IID–non-IID spectrum and complement Dirichlet baselines. A pragmatic protocol is to prefer natural splits when available, include a Dirichlet baseline for controlled ablation, and report heterogeneity diagnostics (e.g., client-size distribution, class coverage, divergence from global labels) for comparability.

Overall, these diverse data-splitting methods not only simulate realistic FL conditions but also reveal limitations in existing algorithms, pointing to a critical need for further investigation. In particular, understanding how each splitting strategy influences convergence behavior, model robustness to adversarial or highly skewed data, and fairness across heterogeneous clients would significantly advance the field. As FL expands into real-world applications, more nuanced data-splitting schemes and rigorous evaluations will be needed to develop robust, scalable, and equitable federated solutions.

\noindent\textbf{RQ4: What are the convergence-privacy trade-offs of common defense methods?}

In FL, most well-designed defenses improve robustness with negligible impact on convergence and clean accuracy. When stronger protection is needed, the cost is typically small (about 1–3\%) clean-accuracy loss with baseline-like convergence, while differential privacy yields certified higher accuracy with bigger trade-offs.
From the Pareto lens on two axes (clean utility and the robustness metric), detection or reweighting and screening approaches usually keep clean utility essentially constant while moving robustness in the right direction; update-shaping or client-side obfuscation brings larger robustness at a small cost. Under non-IID, gains persist but are somewhat attenuated. Notably, high-fidelity replay in unintended exposure attacks enhances retention while expanding the leakage surface. Despite different trade-offs, we recommend designing future methods to explicitly optimize the robustness-utility Pareto frontier, aiming to harmonize convergence and protection rather than trading one off against the other.
\subsection{Open issues}
\label{sec:future}
In FL settings, some additional factors impact the convergence of the global and local models. One of which is the impact of initial model values. Indeed, this is well-stated by Zhu et al. in~\cite{zhu2021federated}, where the authors highlight the effects of non-IID nature or label distribution skews. Such skews result in different initial model states across clients, which can lead to model divergence. This issue becomes particularly pronounced when clients in the FL network have no overlap in labels~\cite{yu2020federated, an2024federated}. For example,  when each client holds data from only a single class, the global model may not converge at all. In such cases, even if local models start with the same initial parameters, they will converge to different ones due to the heterogeneity in local distributions. 

In the original FedAvg paper, McMahan et al.~\cite{mcmahan2017communication} demonstrated through simulations that FedAvg performs better in federated settings when simpler networks are used, as opposed to deeper and more complex architectures. Also, a notable observation from their work is that the use of a deeper network in FedAvg can increase the sensitivity to client data distribution and may lead to non-convergence issues in the case of strong data heterogeneity. This raises the question: \textit{``Can we state that the model reaches a better convergence if it is initialized with a simple and small structure?''}

As a response, Nguyen et al.~\cite{nguyen2022begin} discussed somewhat contrasting by emphasizing the use of pre-trained models as a stable initialization. This significantly optimizes the local models with a higher number of epochs without degrading the final accuracy. At the same time, this avoids performance variability across trials, compared to random initialization when training a model from scratch. Using pre-trained models also enables faster convergence and improved performance, but also helps mitigate non-IID-related issues, as stated in the paper. However, this approach comes with its own limitations. In particular, it assumes the availability of a suitable pre-trained model, which may not always align with the target data distribution. 

From our perspective, while local fine-tuning with pre-trained models can improve convergence, it can increase local computation and communication overhead. Besides, we are concerned that a pre-trained model may also introduce biases if the data used in the pre-training phase differs significantly from the target data. This could potentially impact the generalizability and fairness of the global model. From that, we refine the research question to: \textit{``Can a good model initialization really ensure the model convergence?''}.

To date, numerous approaches have been proposed to address convergence challenges in FL, such as works by Zhou et al.~\cite{zhou2024every}, Wang et al.~\cite{wang2023theoretical}, Bian et al.~\cite{bian2024accelerating}, Feng et al.~\cite{feng2024universally}, Zhai et al.~\cite{zhai2024dog}, and Chen et al.~\cite{chen2020convergence}. However, no official solution has been brought up to create a good model initialization that guarantees convergence. There is also no clear consensus on the previous question of whether simpler initializations or complex, pre-trained weights are more effective, leaving this as an open area for future research. Having a clear statement for this will bring a new, valuable insight for the community regarding setting the initial weights, which is often overlooked in the learning process within federated environments.

\section{Conclusion}
\label{sec:conclusion}
This paper presented a comprehensive analysis of FL from a data perspective, focusing on key aspects such as data traits, data partitioning strategies, data-related vulnerabilities and defense, as well as its impact on convergence behavior.
To bridge the gaps in prior surveys, we link concrete data decisions to optimization outcomes for practitioners, helping them design their system with predictable convergence and stability. We analyze interaction effects among traits that can amplify or cancel one another and turn these findings into clear recommendations for dataset design. Along with that, we provide actionable guidance on choosing the data splitting strategies in experimental environments that more closely reflect real-world heterogeneity, and finally, assess defenses through a robustness-accuracy Pareto lens to present justified trade-offs.
The exploration is particularly significant given that most state-of-the-art FL studies are conducted and evaluated under simulated settings, yet are expected to generalize to real-world deployments.
We also identify underexplored areas and pose open questions to guide future FL surveys.
By formulating and addressing four primary RQs throughout the paper, we provided the first complete guide to understanding and overcoming the data-related challenges in FL, paving the way for more robust and comprehensive learning systems.

\section*{Acknowledgments}
This research was supported by the Research Council of Finland through the 6G Flagship program (grant 318927), Business Finland through the Neural pub/sub research project (diary number 8754/31/2022), University Savoie Mont Blanc through AFREU project, and the Tauno Tönning Foundation Grant for Huong Nguyen and Amirhossein Ghaffari.

\bibliographystyle{elsarticle-num}
\biboptions{sort&compress}
\bibliography{ref}



\end{document}